\documentclass[pre,twocolumn,showpacs,a4paper,superscriptaddress,floatfix]{revtex4-2}
\usepackage{amsmath}
\usepackage{graphicx}
\usepackage{dcolumn}
\usepackage{bm}
\usepackage[monochrome]{xcolor} 
\usepackage{lineno}
\usepackage{lipsum}
\newcommand{\rmd}{\mathrm{d}}
\usepackage{enumitem}

\begin{document}


\title{Spontaneous {\color{red} dust} pulse formation in the afterglow of complex plasmas under microgravity conditions}

\author{M. Chaudhuri$^*$}
\affiliation{Max-Planck-Institut f\"ur extraterrestrische Physik (MPE), D-85741 Garching, Germany}
\altaffiliation[Present address:~]{ASML, Netherlands}
\thanks{M.C. initiated the project investigations during his work period at MPE, Garching, Germany. L.C., E.T. and M.S. completed the draft in its present form.}
\author{L. Cou\"edel}
\affiliation{Physics and Engineering Physics Department, University of Saskatchewan, 
Saskatoon, SK S7N 5E2, Canada}
\affiliation{CNRS, Aix-Marseille Universit\'e, Laboratoire PIIM, UMR 7345, 13397 Marseille, France}
\author{E. Thomas, Jr.}
\affiliation{Physics Department, Auburn University, Auburn, Alabama 36849, USA}
\author{P.~Huber}
\affiliation{Institut f\"ur Materialphysik im Weltraum, Deutsches Zentrum f\"{u}r Luft- und Raumfahrt (DLR), 82234 We{\ss}ling, Germany}
\author{A.\,M.~Lipaev}
\affiliation{Joint Institute of High Temperatures, Russian Academy of Sciences, 125412 Moscow, Russia}
\author{H.\,M.~Thomas}
\affiliation{Institut f\"ur Materialphysik im Weltraum, Deutsches Zentrum f\"{u}r Luft- und Raumfahrt (DLR), 82234 We{\ss}ling, Germany}
\author{M. Schwabe}
\affiliation{Institut f\"ur Materialphysik im Weltraum, Deutsches Zentrum f\"{u}r Luft- und Raumfahrt (DLR), 82234 We{\ss}ling, Germany}

\begin{abstract}
A new type of nonlinear {\color{red} dust pulse} structures has been observed in afterglow complex plasma under microgravity condition on board the International Space Station {\color{red}(ISS)}. The {\color{red} dust} pulses are triggered spontaneously as the plasma is switched off and the particles {\color{red} start to} flow through each other (uni-directional or counter-streaming) in the presence of a low-frequency external electric excitation. The pulses are oblique with respect to the microparticle cloud and appear to be symmetric with respect to the central axis. A possible explanation {\color{red} of this observation} with the spontaneous development of a double layer in the afterglow {\color{red} of complex} plasma is described.  

\end{abstract}


\maketitle

\section{Introduction}


A complex (dusty) plasma is an overall charge neutral assembly of ions, electrons, highly charged microparticles ({\color{red}dust/}grains) and neutral gas. The microparticles are negatively charged due to their interactions with the surrounding free electrons and ions of the plasma {\color{blue}and interact with each other, in analogy to atoms in conventional liquids and solids \cite{Morfill2009}. They are} large enough to be visualized individually and, hence, their motion can be easily tracked (with the help of high-speed high resolution video cameras). The overall dynamical time scales associated with the dust component are in the range 10-100~Hz. These unique features allow experimental investigations with high temporal and spatial resolution using relatively simple video microscopy methods. 
The laboratory research on complex plasmas received considerable attention after the discovery of plasma crystals~\cite{ThomasPRL1994,Chu1994PRL,Melzer1994,Hayashi_1994,ThomasNATURE}. Complex plasmas are used as model systems to investigate various phenomena (e.g, phase transitions, self-organization, waves, transport, etc.) at the most fundamental kinetic level~\cite{Fortov2005,Shukla2002,Vladimirov2005,Morfill2009,Chaudhuri2010}. Apart 
from their relevance in fundamental studies, dusty plasmas also play an important role in connection to astrophysical plasmas~\cite{Mendis2013}, planetary rings~\cite{Morfill2009}, technological plasma applications~\cite{Vladimirov2005,Bouchoule1999}, fusion devices~\cite{Winter2004,Fortuna-Zalesna2017,Ratynskaia2017}, etc. 

\begin{figure}[t]
    \includegraphics[width=\linewidth]{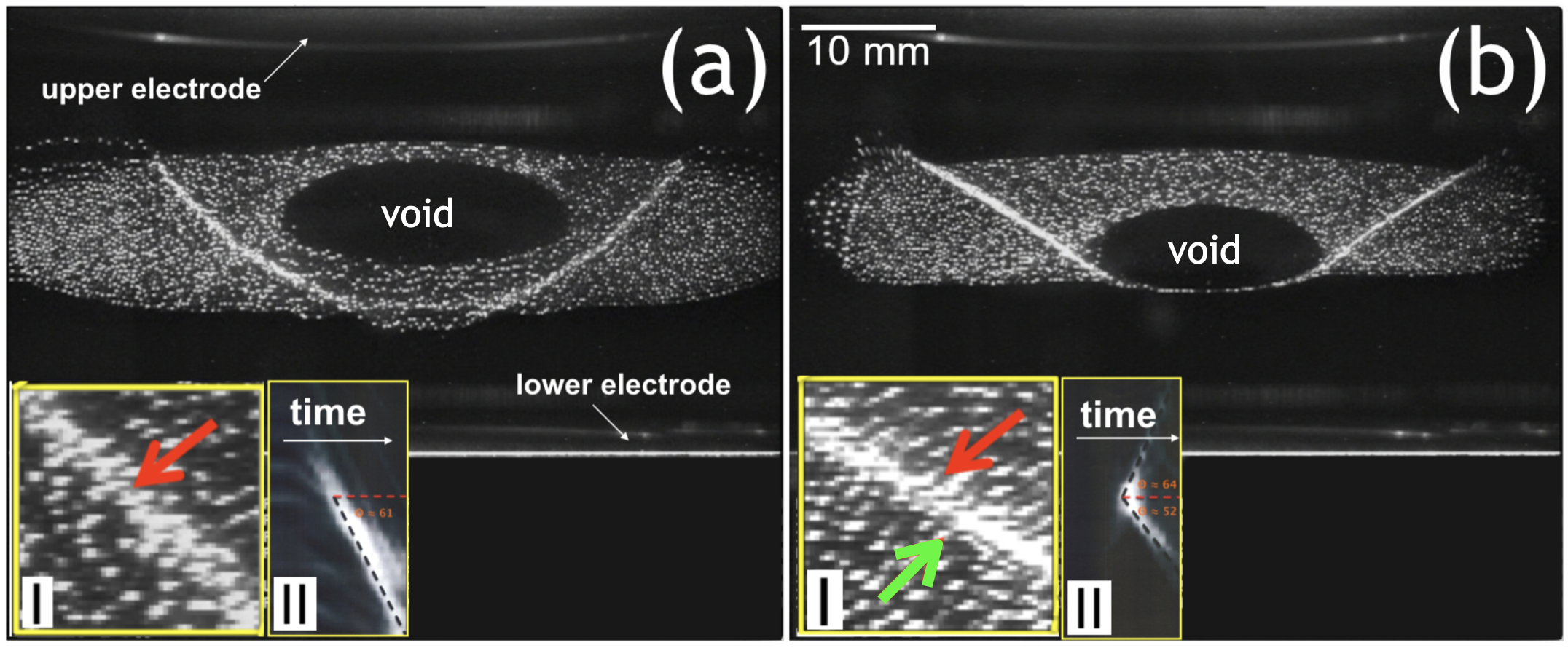}
    \caption{\label{fig:observation} The nonlinear oblique pulse generation in an afterglow plasma as observed by the overview camera{\color{blue}. The upper and lower electrodes as well as the central particle-free void are visible. Microparticle pulses are discernible as bright lines, corresponding to higher dust density, which are} symmetric w.r.t the vertical mid-line of the plasma chamber. These structures are associated with the inter-penetration of particles in (a) one direction (shown by red arrow) 
    or (b) in both directions 
    (shown by red and green arrows) in the insets (I) (fields of view $3.5 \times 3.5 $~mm$^2$). This directional flow feature is also evident from the periodgram{\color{blue}s (time-space-plots)} with one arm (uni-direction) and two arms (bi-direction) as shown in the insets (II) {\color{blue}(field of view: {\color{red}2}~mm $\times$ {\color{red}320}~ms)}}
\end{figure}

Experiments on complex plasmas are often performed under microgravity conditions on board the International Space Station (ISS) or in parabolic flights \cite{FortovJETP1998,Morfill1999a,Nefedov2003,Klindworth2004,ThomasNJP,Himpel2011,Takahashi2014,Pustylnik2016}, since the presence of the {\color{red}gravitational force in ground based laboratories strongly affects the microparticle dynamics} \cite{Melzer2019}. Under those conditions, in a capacitively coupled complex plasma, usually a microparticle-free central region appears, the so-called \textit{void}. The void formation is a result of the interplay of the drag force from ions streaming towards the plasma chamber walls and the electric force confining the microparticles \cite{Goree1999,Khrapak2002,Kretschmer2005,Lipaev2007}. Despite this central structure, compared to experiments on ground, weightless microparticle clouds form more homogeneous three dimensional (3D) 
structures. This facilitates the study of propagating features such as shocks \cite{Samsonov2003}. {\color{blue}Here, we discuss the formation of microparticle pulses as shown in Fig.~\ref{fig:observation} in an afterglow complex plasma. These pulses formed when the plasma turned off under microgravity conditions in the presence of a direct current (DC) electric field and could consist of either uni-directional (Fig.~\ref{fig:observation}(a)) or bi-directional (Fig.~\ref{fig:observation}(b)) high-density pulses of propagating microparticles.}

{\color{violet}{In any complex plasma experiment, a key parameters is the electric charge carried by the microparticles. Indeed, the electric charge determines the strength of the (screened) Coulomb force between the microparticles and is also an important parameter in other forces such as the ion drag force. One should nevertheless keep in mind that the net electric charge carried by the microparticles is not fixed but is determined mostly by the fluxes of electrons and ions coming from the surrounding plasma. In that sense there is a strong coupling between the local plasma parameters and the microparticle charge. This effect also carries on in the post-discharge phase (afterglow). 
		
Residual electric charges of microparticles in the afterglow were observed for the first time under microgravity conditions by Ivlev et al. \cite{Ivlev2003}. Systematic investigations by Cou{\"e}del et al. using nanoparticles grown directly inside a capacitively-coupled RF discharge show that decharging processes and residual charges are strongly linked to the diffusion of electrons and ions in the post discharge 
\cite{Couedel2006,Couedel2008,Couedel2008b,Couedel2009a,Layden2011a}. Theoretical investigations by Denysenko et al. also reveal that the presence of metastable atoms can considerably influence the residual electric charges of the microparticles \cite{Denysenko2011,Denysenko2013}. 
Extensive experiments on microparticle decharging using the PK-3 Plus laboratory demonstrate that the presence of an electric external field in the post discharge can shift the microparticle charge distribution from globally neutral (equal number of positively and negatively charged particles with a sharp peak at 0 charge) to globally positive \cite{Woerner2013}. Finally, recent experiments by Meyer and Merlino show that the microparticle cloud motion is very sensitive to the initial cloud configuration and shape and can go from a simple Coulomb expansion to more complex behavior such as Coulomb fission \cite{Meyer2016,Merlino2016}. All of the afterglow studies suggest that the residual electric charges and the associated dynamics of the microparticles in the afterglow phase keep an imprint of the plasma conditions at the moment the discharge is switched off.


The goal of this article is to explore a new phenomenon in afterglow complex plasmas under microgravity conditions. As previously mentioned, when a plasma with embedded microparticles extinguishes, the microparticles inside the system remain electrically charged for a certain period of time, and their motion depends directly on the plasma condition at the moment the discharge is switched off (external electric field, dust cloud configuration and density, metastable atoms, etc) 
\cite{Ivlev2003,Couedel2006,Couedel2008,Couedel2008b,Couedel2009a,Layden2011a,Denysenko2011,Denysenko2013,Woerner2013,Meyer2016}. 

In the present article,  we report on the generation of uni-/bi-directional pulse structures in the plasma afterglow reminding of the structure of previously observed oblique waves in complex plasma clouds under microgravity conditions \cite{Piel2006,Schwabe2008}. These pulse
structures were observed in the particle cloud adjacent to the void immediately after the plasma switched off while a low frequency electric signal was applied using a function generator in addition to the radio-frequency (RF) field from the RF generator. 

The article is organized as follows: Section 2 briefly presents the PK-3 Plus laboratory and the experimental parameters. Section 3 describes our investigation of the phenomena using different data analysis techniques. Section 4 presents a simple molecular dynamics simulation taking into account the spatial distribution of residual charges to explain the observed phenomena. Finally, the discussion and conclusion are presented in Section 5.


\section{Experimental setup and observations}

\begin{figure}
	\includegraphics[width=\linewidth]{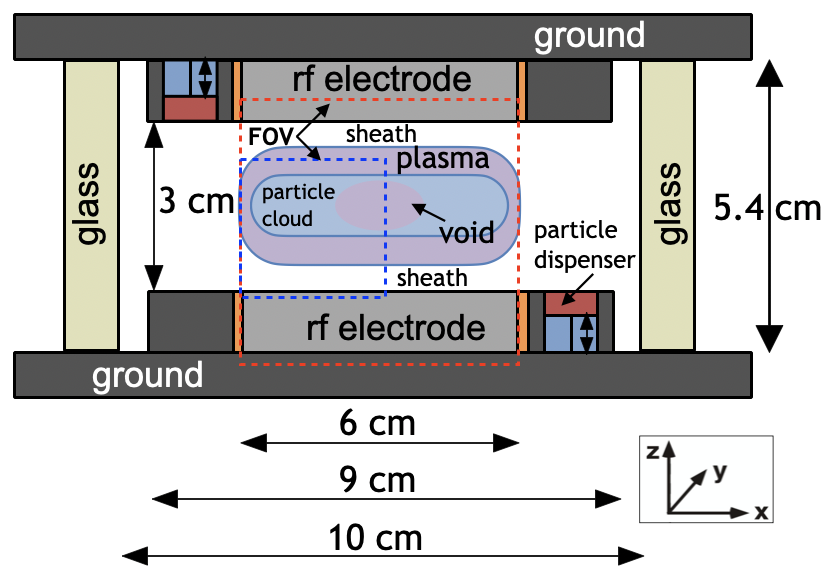}
	\caption{\label{fig:setup}Sketch of the PK-3 Plus chamber seen from the side~\cite{ThomasNJP}. The fields of view (FOV) of the overview (red dashed line) and quadrant view (blue dashed line) cameras are shown which were used for the analysis of our experiments: Fig.~\ref{fig:observation} (overview) and Figs.~\ref{fig:ptv},~\ref{fig:piv},~\ref{fig:flow} (quadrant view). Typically a particle free region (the so-called ``void'') formed almost at the center of the particle cloud under microgravity conditions.}
\end{figure}

The experiments presented here were performed in the PK-3 Plus Laboratory \cite{ThomasNJP,Khrapak2016,Schwabe2018,Thomas2019} which was a Russian-German laboratory hosted on board the ISS during 2006 -- 2013. This microgravity laboratory made it possible to explore the interdisciplinary features of complex plasmas with several dedicated experiments on diverse topics such as the heartbeat instability and its effect on the microparticle dynamics \cite{Heidemann2011,Schwabe2017}, phase separation \cite{Khrapak2012b}, lane formation~\cite{SutterlinPRL}, crystallization \cite{Khrapak2012b,Molotkov2017,Naumkin2018,Zhukhovitskii2019}, electrorheological plasmas \cite{Ivlev2008}, and interface effects \cite{Yang2017,Sun2018}. 

A sketch of the PK-3 Plus chamber is shown in Fig.~\ref{fig:setup}. Its heart consisted of an RF plasma chamber where the electrodes of 6~cm diameter were separated by 3~cm and were driven in push-pull mode. Microparticles of various sizes could be injected into the plasma with dispensers mounted in guard rings surrounding the electrodes. A slice of the resulting microparticle cloud was illuminated with a laser, and the light scattered by the microparticles was recorded with cameras, the fields of view of which are shown in Fig.~\ref{fig:setup}. 

In addition to the RF electric field, it was possible to apply another electric signal to the electrodes with a function generator (FG) alternating at frequencies up to 255~Hz. This frequency is very small with respect to the plasma frequencies of the electrons and ions, so that they were fully able to follow the field. The function generator could be used to, for instance, excite waves \cite{Schwabe2008} or to drive the motion of charged microparticles in the plasma afterglow \cite{Woerner2013}. In the experiments discussed here, the plasma switched off while such a low-frequency electric field was applied, which was essential to excite the observed pulse. When the plasma extinguished in the absence of an electric field from the FG, the particles did not flow through each other and hence no such pulse generated.

Typically, when the plasma extinguishes under microgravity conditions, the microparticles remain more or less stationary at their positions. There is often a slow drift of the particles following a small temperature gradient in the system due to differential heating of the sides of the chamber by the electronical systems. 

If an electric field is applied after the plasma is switched off, the charges of the microparticles can be measured by analyzing their motion following the electric field. W\"orner et al. \cite{Woerner2013} find that under those conditions, the particle charge distribution in the afterglow peaks at zero with a homogeneous spatial distribution of the charges. However, the authors show that if an electric field is applied with the function generator at the moment that the plasma turns off, the particle charge strongly depends on the positions of the particles at that time, and the charge distribution shifts towards positive values. 

In contrast to the experiments discussed above \cite{Woerner2013}, in our experiments dust pulses propagated through the microparticle cloud after the plasma turned off while an electric field was present (that is, under similar conditions when the position dependent charges where found in previous experiments - however, since the experiments presented here were not dedicated decharging experiments, the discharge was not turned of on purpose, but extinguished on its own). 
These pulses appeared to be oblique w.r.t. the horizontal mid plane of the chamber in the range 25$^{\rm o}$ to 45$^{\rm o}$ and also, they were symmetric w.r.t. the void as shown in Fig.~\ref{fig:observation}.

\begin{figure}[ht!]
	\includegraphics[width=0.9\linewidth]{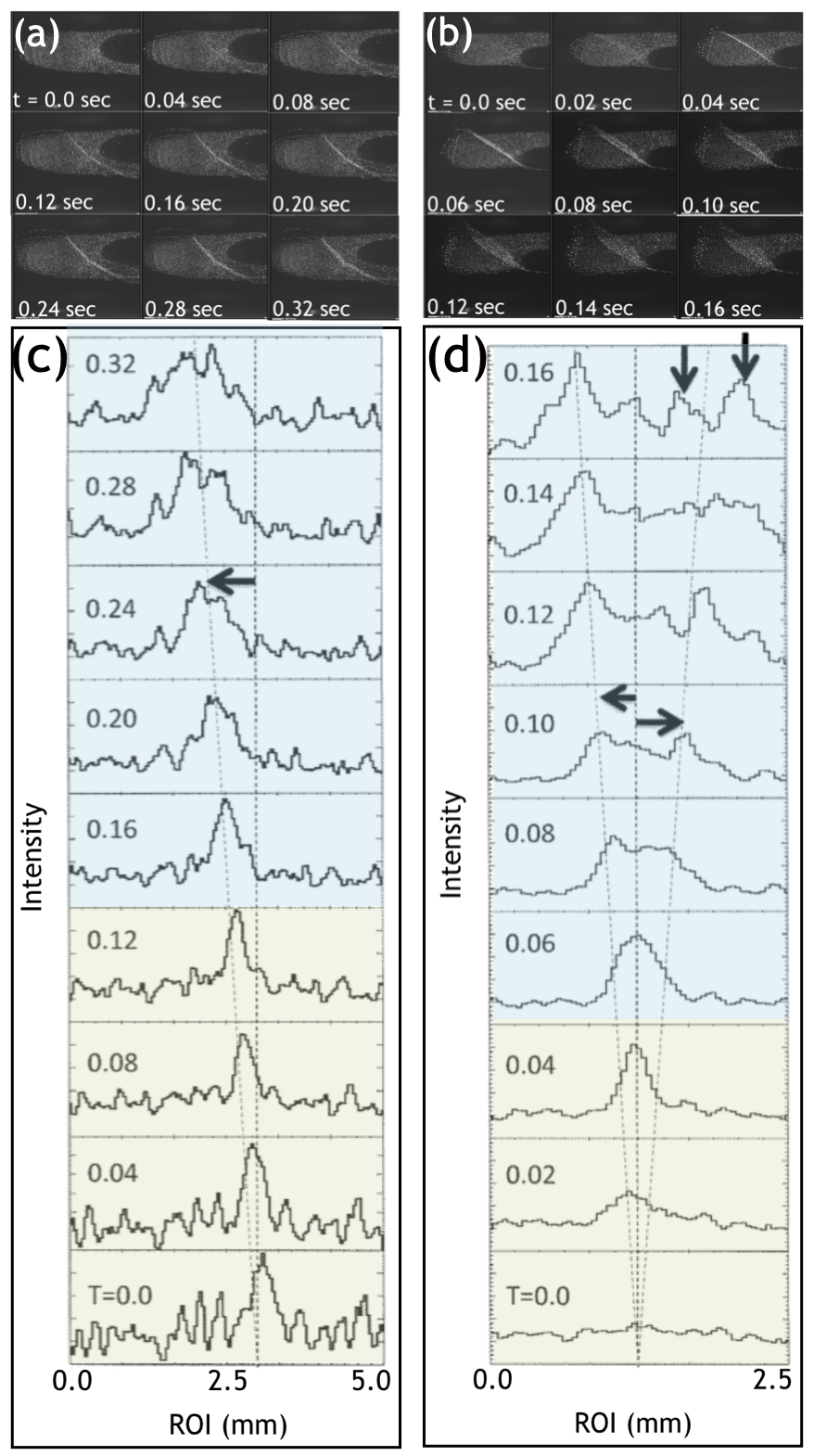}
	\caption{\label{fig:flow} Time evolution of (a) uni-directional (Exp.~I in Tab.\ref{tab:table1}) and (b) bi-directional pulse (Exp.~III) propagation. The amplitude and width evolution of both types of pulses are shown in (c) and (d) for various instances of time since plasma off (time stamps in s). In either case, initially the amplitude of the pulse increased and the width decreased when the particles approached each other (approaching phase) as shown with a green background. After reaching the maximum pulse height, the pulses started dispersing with decreasing amplitude and increasing width over time (dispersive phase) as shown with a blue background. It is clear that in (c) the pulse dispersed in one direction whereas in (d) it occurred in both directions. These features are also evident from the periodgram profile as shown in the insets-II within Fig.~\ref{fig:observation} (one arm for uni-directional pulse and two arms for bi-directional pulse). They also created multiple peaks in the later stage as shown by vertical arrows.}
\end{figure}

The parameters of the experiments performed in the PK-3 Plus Laboratory on board the ISS during various missions in which oblique dust pulses were observed in an afterglow plasma are shown in Table~\ref{tab:table1}.

\begin{table*}
	\caption{\label{tab:table1}
		Data for different experiments (Exp.): particle diameter ($\phi$), power $P$, pressure $p$, Epstein damping coefficient $\gamma_{\rm Ep}$, amplitude (A$_{\rm FG}$) and frequency (f$_{\rm FG}$) of the excitation generated by the function generator (FG), electric field $E_{\rm AC}$, velocity of the approaching $v_{\rm a}$ and dispersing pulse $v_{\rm d}$. Two types of pulses were observed: uni-directional (particles move in one way) and bi-directional (particles move in both ways). The uncertainty in the velocity measurements is $\sim$12--35\%. The major contribution to the large uncertainty for some experiments is due to poor fitting results at the larger time scale of fast pulse dispersion (see text for detailed procedure).}
	\centering
%
%
\begin{ruledtabular}
	\begin{tabular}{|cccccccccccccc|}
		
		Exp. & Mis- & Gas & Particle & $\phi$ & $P$ & $p$ & $\gamma_{\rm Ep}$ & $A_{\rm FG}$ & $f_{\rm FG}$  & $E_{\rm AC}$ & $v_{\rm a}$ & $v_{\rm d}$ & pulse\\
		& sion &  & type &  ($\mu$m)  & (mW) & (Pa) & (s$^{-1}$) & (V) & (Hz)  & (V/cm) & (mm/s) & (mm/s) & type\\\hline
		I & 03 & Neon & MF & 9.19 & 480  & 15.4 & 14.5 & 13.3 & \:48 & \:8.9 & \:3.3$\pm$0.5 & \:3.2$\pm$0.3 & uni-dir \\\hline
		II & 15 & Neon & Silica & 1.55  & 770  & 15.3 & 70.3 & 24.9 & 255 & 16.6 & \:2.5$\pm$0.3 & \:2.3$\pm$0.4 & uni-dir\\\hline
		III & 04 & Argon & MF & 6.81  & \:74  & \:9.9 & 17.7 & 13.3 & 100 & \:8.9 & 23.4$\pm$2.6 & 13.6$\pm$0.9 & bi-dir\\\hline
		IV & 08 & Neon & MF & 6.81  & 310 & 30.3 & 38.6 & 28.3 & 100 & 18.9 & \:4.6$\pm$0.6 & \:8.5$\pm$0.6 & bi-dir\\\hline
		V &20 & Argon & Silica & 1.55  & \:61 & \:9.4 & 60.5 & 13.3 & 255 & \:8.9 & 26.2$\pm$4.4 & 17.4$\pm$1.6 & bi-dir\\\hline
		VI & 20 & Argon & MF & 2.55 & \:61 & \:9.4 & 45.0 & 13.3 & 255 & \:8.9 & 30.5$\pm$4.8 & \:9.7$\pm$1.2 & bi-dir\\
		
	\end{tabular}
\end{ruledtabular}
\end{table*}

The pulses formed when the particles flowed through each other in one direction or in both directions. In either case, two propagation phases can be observed: (1) an {\it approaching} phase when the pulse amplitude increased with decreasing pulse width over time, and (2) a {\it dispersive} phase when the amplitude decreased with increasing pulse width over time as shown in Fig.~\ref{fig:flow} and Fig.~\ref{fig:shock-fronts}} (see below for a more detailed description). It is to be noted that the pulse propagation strongly depended on the gas pressure: the pulses moved faster in low gas pressure due to the decreased damping by neutral particles.

\section{Data analysis}

The nonlinear pulse generation and its time evolution are shown in Fig.~\ref{fig:flow}. Initially the unperturbed microparticle cloud contained a void (dust free region), which is visible in the right of the images. As the plasma switched off, the discontinuous microparticle pulse was triggered locally in a narrow region within the microparticle cloud. Then it evolved very fast as a result of flows of particles from both of its sides without affecting the void structure. It quickly lost its high-density structure and almost completely dissipated within 0.3~s for the fast bi-directional pulse and within 0.6~s for the uni-directional pulse. Note that the damping times, calculated as the inverse of the Epstein damping frequencies \cite{PhysRev.23.710.epstein,Jung2015}, were 0.06~s and 0.07~s, respectively.

Next, in order to measure the velocity of the pulses, we define a region of interest (ROI) that is aligned with the direction of propagation of the pulses. We then calculate periodgrams (space-time-plots) \cite{Schwabe2007} by averaging the pixel intensities in the ROI along the direction parallel to the pulse propagation direction. The resulting intensity plots are shown in the bottom panels in Fig.~\ref{fig:flow} for several instances of time. In the experiment shown  in the left column, a single pulse propagated, whereas in that in the right, two pulses moved in opposite directions. This is indicated with the dashed lines in Fig.~\ref{fig:flow}} along with the arrows. 


\begin{figure}
	\includegraphics[width=0.85\linewidth]{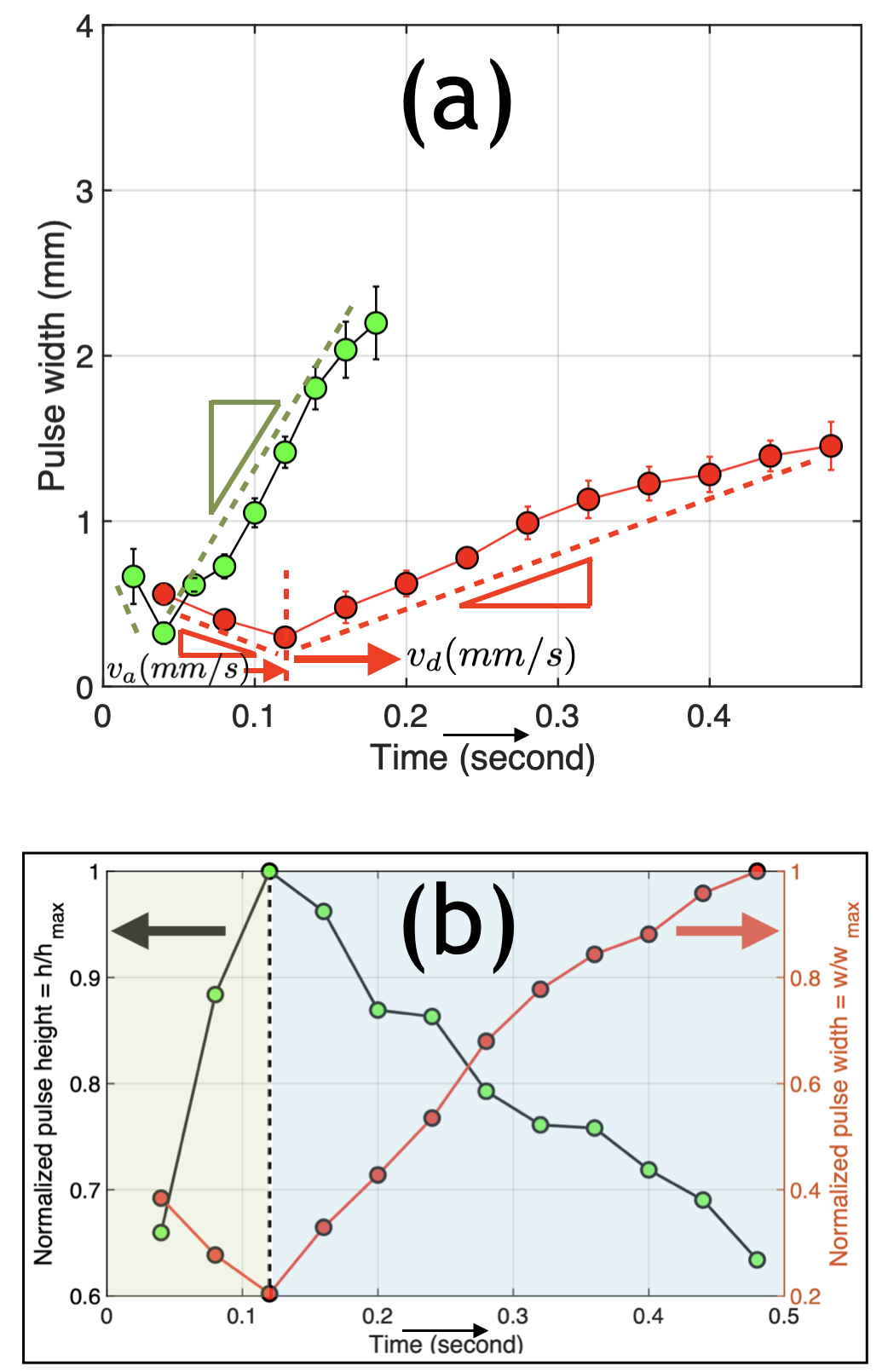}
	\caption{\label{fig:shock-fronts}{\color{red} (a) Time evolution of the pulse widths for Exp.~I  (uni-directional pulse propagation) as marked by the red circles and Exp.~III (bi-directional pulse propagation) as marked by the green circles. It is clear that the bi-directional pulse propagated faster than the uni-directional pulse. (b) The minimum of the normalized pulse width (left y-axis) corresponds to the maximum of normalized pulse height (right y-axis) at the transition point for Exp.~I: before this transition point the particles approached each other with the velocity $v_a$ (approaching phase with green background) and after this time the particles dispersed with the velocity $v_d$ (dispersive phase with blue background).}}
\end{figure}

The intensity plots shown in Fig.~\ref{fig:flow} also make it possible to measure the evolution of the pulse widths as a function of time. The results for the same experiments given above are shown in Fig.~\ref{fig:shock-fronts}. It can be seen that after an initial steepening, the pulses quickly widened and eventually dissipated. In this respect, we fitted the pulse shape in each time frame with a Gaussian distribution, $f(x) = c_0\exp{(-z^2/2)}$ where $z = (x - c_1)/c_2$ and $c_0, c_1, c_2$ are fitting constants corresponding to height, center and width (the standard deviation) of the Gaussian. The full width at half maximum (FWHM) of the Gaussian was computed as $2\sqrt{2\log(2)} c_2$ and is considered as the pulse width. Figure~\ref{fig:shock-fronts} shows the variation of this pulse width with time. The pulse width velocity for the uni-directional pulse is much lower than that for bi-directional pulse.
However, this difference in pulse speed cannot be explained by the difference in gas pressure alone, since the damping coefficient was nearly identical for the two experiments. 


\begin{figure}
	\includegraphics[width=\linewidth]{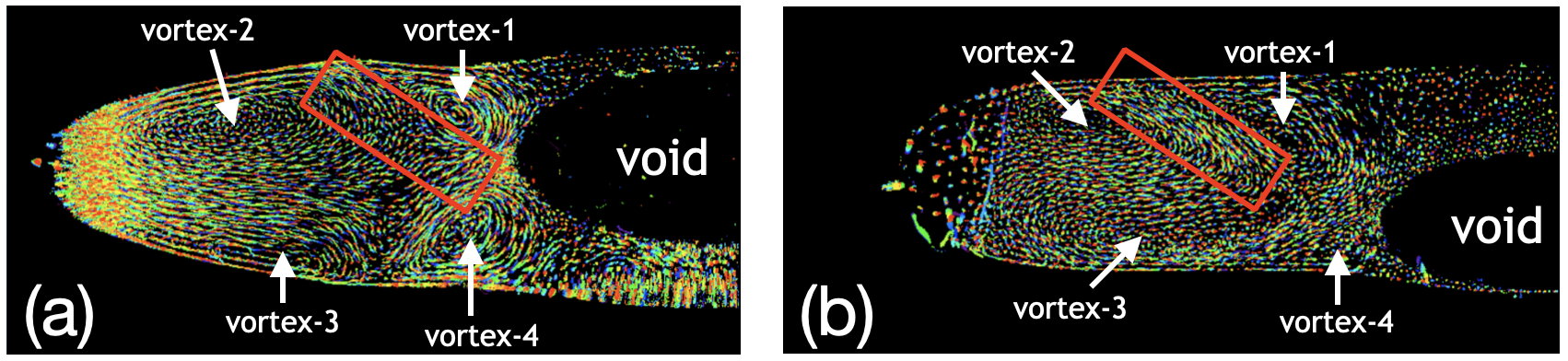}
	\caption{\label{fig:ptv}Superposition of multiple color-coded images of the microparticle motion before the plasma switched off, resulting in (a) a uni-directional pulse (Exp-I) and (b) a bi-directional pulse (Exp-III). The pulse forming 'depleted' zone is marked by the red box. For the uni-directional pulse, four strong vortices were present in the particle cloud, and the particles in the red box had clear x-directional velocities, in particular at the intersection of vortex 1 and 4. But for the bi-directional pulse, the particles in the red box moved in a 'correlated, directed and oblique' way. In this case, the vortex motion was not as strong as for uni-directional pulse.} 
\end{figure}

To elucidate the difference between the situations leading to uni-directions or bi-directional pulse formation, we consider the state of the microparticle cloud before the plasma switched off. Fig.~\ref{fig:ptv} shows superpositions of particle positions color-coded to indicate time. Before the plasma switched off, the particles performed the vortex motion typical for weightless complex plasmas in RF chambers \cite{Fortov2003,Goedheer2003,Schwabe2014,Bockwoldt2014} (see the Appendix \ref{sec:chargeforce} for the charge and force estimation before plasma off). Four vortices were observed with different sizes at different locations. Their interactions were unique and distinct for uni- and bi-directional pulses. Within the pulse forming region as shown by the red rectangle in Fig.~\ref{fig:ptv}, the vortices interacted in opposite directions for the uni-directional pulse, thus forming a depleted region with no correlated directed motion of particles. However, for the bi-directional pulse, the vortices interacted in such a way that a clear correlated motion of all particles in the pulse forming region from the center to the upper boundary could be observed.

\begin{figure}
    \includegraphics[width=\linewidth]{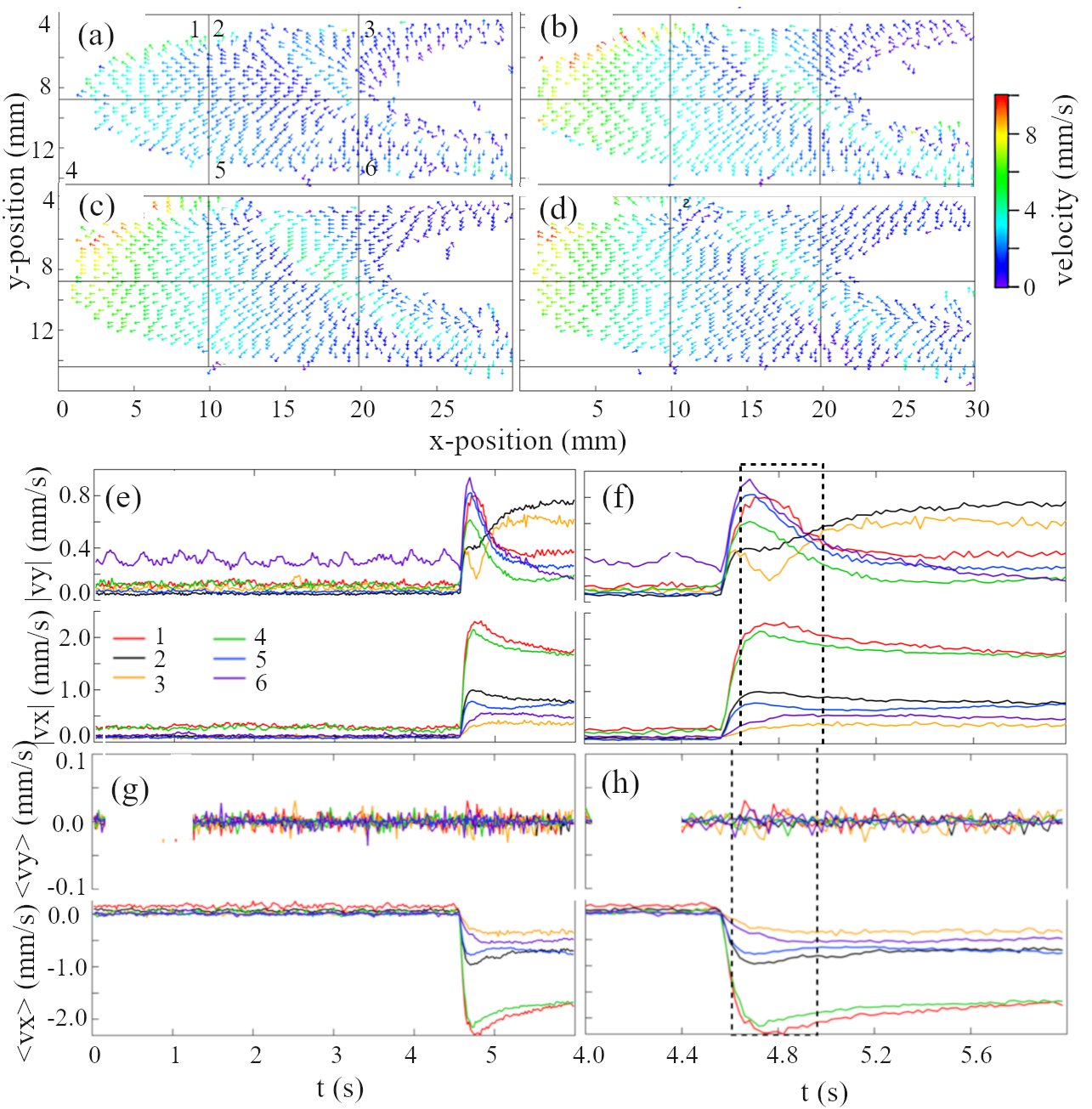}
    \caption{\label{fig:piv}{\color{blue} (a)-(d) PIV analysis for the uni-directional pulse in Exp.~I at different times after the plasma switched off. The time step between panels (a)-(d) is 0.1~s. For display purposes, only every third vector is shown. All vectors are drawn with the same length, and their color represents their velocities. The field of view is divided into 6 regions, indicated by solid lines. (e)-(h) Absolute and average horizontal and vertical velocities in the regions defined above as a function of time.}}
\end{figure}

In order to study the motion of the microparticles in more detail, we performed particle image velocimetry (PIV) on the experiment with the uni-directional pulse (Exp.~I). In PIV, a regularly spaced two-dimensional grid is placed over the images, and cross-correlation is
performed between two successive images \cite{ThomasJr1999} to reconstruct the spatial velocity profile. Fig.~\ref{fig:piv} shows the velocities in the cloud for four instances of time after the plasma switched off, as well as the temporal evolution of the magnitude and averages of the velocity components.

\begin{figure}
	\includegraphics[width=\linewidth]{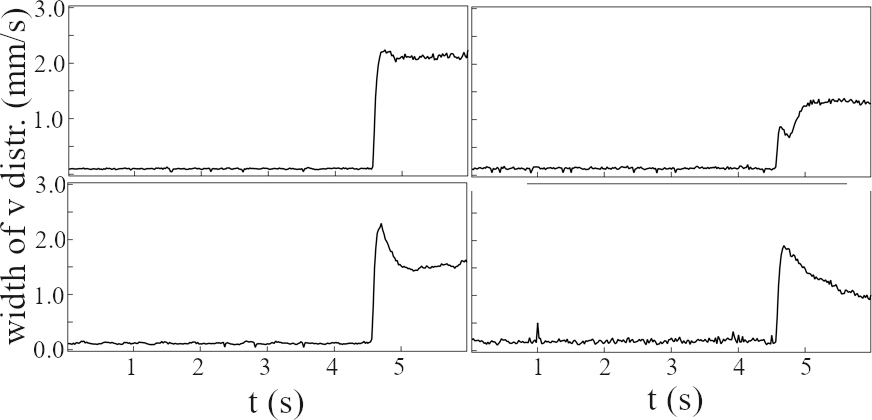} 
	\caption{\label{fig:pivwidth}{\color{orange} Time evolution of the width of the PIV velocity distributions for Exp.~I for Regions 2, 3, 5 and 6 (corresponding to Fig.~\ref{fig:piv}). In Regions 1 and 4, the leftmost regions of the images, we were unable to obtain reliable fits to the distribution function. }}
\end{figure}

It can be seen that, once the plasma was off, the cloud expanded and the particles moved to the sides. The formation of the pulse is clearly visible as a green-yellow region to the left of the void. It is also important to note the bifurcation in the particle motion (e.g., both upward and downward moving particles) in different regions of the particle cloud.  This behavior is discussed and characterized in Figs.~\ref{fig:pivwidth} and~\ref{fig:flow}.

We further divided the cloud into six panels, shown as black solid lines in Fig.~\ref{fig:piv}(a)-(d), and plot the magnitude of the vertical ($y$) and horizontal ($x$) velocities as well as their averages as function of time in Fig.~\ref{fig:piv}(e)-(h). At the moment of plasma off, at $t \approx 4.5$~s, the absolute values of the velocities and that mean horizontal velocity, $<vx>$, rapidly increase. The bifurcated motion of the particles in the different regions is well visible, especially in the absolute values of the vertical velocities, $|vy|$.

In addition to the analysis of the particle velocities, it is also useful to examine the associated distribution of the particle velocities in the cloud and in the six regions. Here, the magnitude of each of the PIV velocity vectors was determined and then binned in steps of $\Delta v = 0.1$~mm/s.  The resulting distribution was then fit using a two-dimensional speed distribution that is given by:  
\begin{equation}
        f(v) = A \frac{v}{\sigma} exp(-\frac{v^2}{\sigma ^2}).
   \end{equation}
We chose this approach over the usual 1-dimensional, directional fits because of the bifurcated velocity motion mentioned above.

Figure~\ref{fig:pivwidth} shows the resulting widths of the velocity distributions for Regions 1, 2, 5 and 6. At the moment of the plasma switch-off, there was a rapid increase in the width of the velocity distribution as the particles underwent a rapid expansion. This appears in Fig.~\ref{fig:pivwidth} as a significant broadening of the distribution, but it is uncertain whether this should be identified as a heating process. It is important to note that this broadening took place on a much faster timescale than the formation of the shock, which further suggests that a mechanism other than shock heating was occurring in the system.

Further, we performed MD simulations to shed light on the formation mechanism of the pulses. 

\section{Simulations}

The simulations were performed in two steps.

1) Since the number of microparticles in the PK-3 Plus chamber was very large (tens of thousand), it was not practical to simulate the entire 3D cloud of microparticles. For this reason, a few thousand of charged microparticles were confined in the simulation box using parabolic confinement due to the ambipolar electric field in the $x$- and $z$- directions, and the $y$-direction used periodic boundary conditions. For each particle, the motion was described by the following equation:
 \begin{align}
        m_\rmd \frac{\rmd \vec{r}_i}{\rmd t} = & \sum_{j \neq i} \vec{F}_{ij}  - m_\rmd \nu \frac{\rmd \vec{r}}{\rmd t} 
          - m_\rmd \omega^2_\mathrm{conf,x} \vec{x}_i\nonumber \\
         &- m_\rmd \omega^2_\mathrm{conf,z} \vec{z}_i + L(t),
\end{align}

where $\vec{r}_i$ is the position of the $i^{\mathrm{th}}$ microparticle, $m_\rmd$ the mass of the microparticles, $\nu$ the neutral drag coefficient and $\omega^2_\mathrm{conf, x (y)}$  the confining angular frequencies. Assuming screened coulomb interactions between the microparticles, the interparticle force was 

\begin{equation}
        \vec{F}_{ij} = \frac{Q_\rmd^2}{4\pi \epsilon_0 r_{ij}^3}\exp \left(-\frac{r_{ij}}{\lambda_\mathrm{D}} \right) \left( 1+
        \frac{r_{ij}}{\lambda_\mathrm{D}}\right)\vec{r}_{ij},
\end{equation}
with $\vec{r}_{ij}=\vec{r}_i - \vec{r}_j$, $r_{ij}= |\vec{r}_{ij}|$, $Q_\rmd$ denoting the microparticle charge, $\lambda_\mathrm{D}$ the screening length and $\epsilon_0$ the permitivity of vacuum. $L(t)=\sqrt{2\nu k_\mathrm{B} T}$ is the Langevin force used to maintain the cloud at temperature $T$. 

\begin{figure}[t]
	\centering
	\includegraphics[width=0.95\columnwidth]{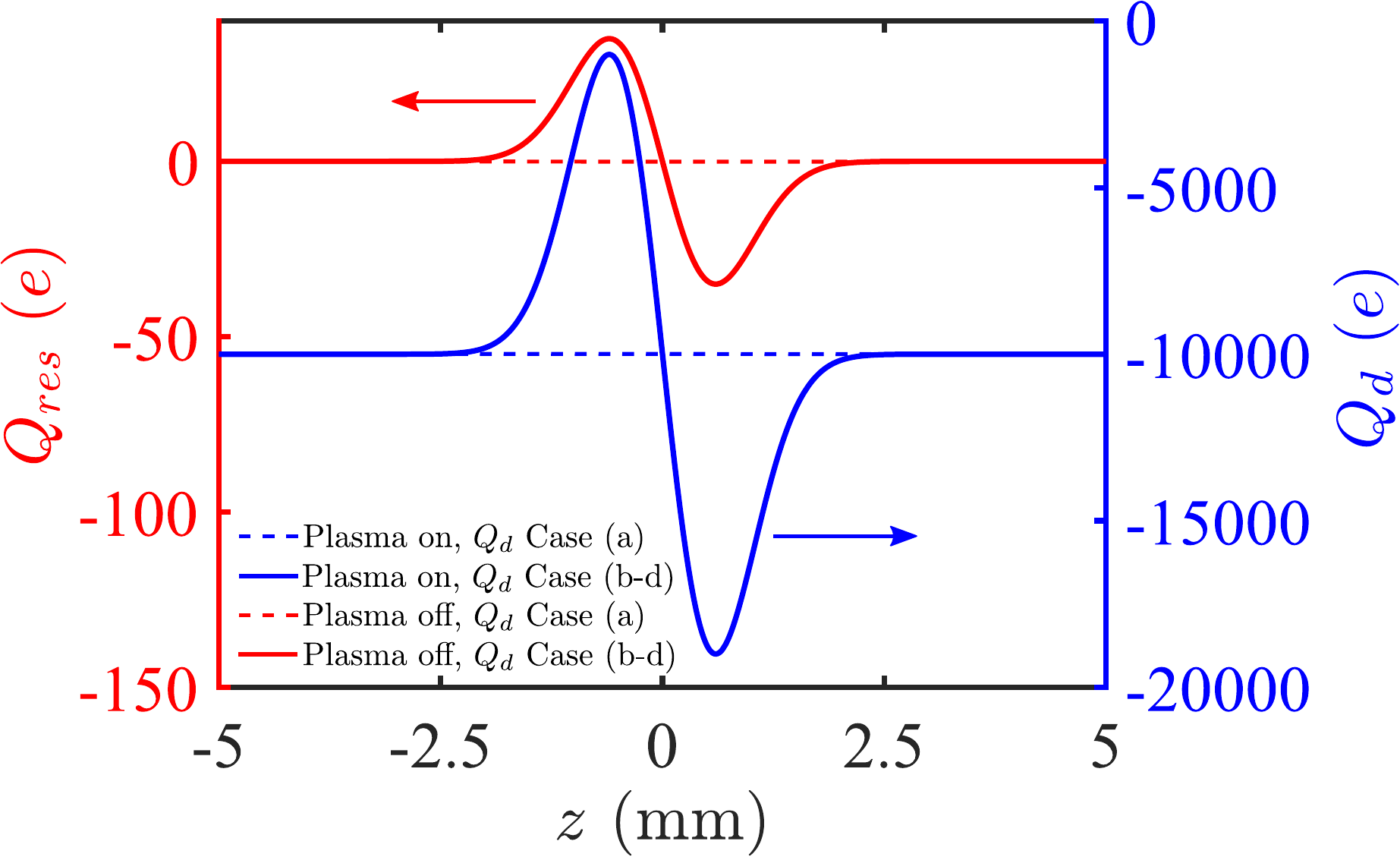} 
	\caption{Dust particle charges as a function of their position at plasma extinction (blue curves) and after decharging 
		(red) curves. The decharging time was set to $\tau_\mathrm{D}=1\ \mathrm{ms}$.}
	\label{fig:charge_distribution}
\end{figure}

Before solving the system of equations, all quantities were normalized. The energy unit was defined as $E=Q_\rmd^2/(4\pi \epsilon_0 \lambda_\mathrm{D})$, the distances were normalized with respect to $\lambda_\mathrm{D}$, and the masses were normalized with respect to $m_\rmd$. Time and frequency were normalized using the time normalization constant $\tau=\sqrt{m_\rmd\lambda_\mathrm{D}^2/E}$. The time step was $\rmd \tau=0.005\cdot \tau$. The temperature was adjusted in order to obtain a fluid microparticle cloud. Starting from random positions, the simulation was run for $\sim1,000,000$ time steps (Beaman algorithm with predictor-corrector scheme) until a stable cloud was obtained. Microparticle cloud parameters similar to the experimental ones were chosen: $Q_\rmd=-10,000e$, $m_d=2.5\cdot 10^{-13}$~kg, $\omega_x=0.05\ \mathrm{s^{-1}}$, $\omega_z=0.15\ \mathrm{s^{-1}}$, $\lambda_\mathrm{D}=120\ \mathrm{\mu m}$, $T=500$~K, and $\nu=10\ \mathrm{s^{-1}}$. 

The aim of this equilibration run in the ``plasma-on phase'' was to obtained a realistic distribution of the microparticles in the interelectrode space, not to simulate the complex plasma discharge itself. Thus, possible spatial inhomogeneities of dust particle charges and electric fields due to the existence of a double layer are not taken into account but could be added afterwards when simulating the afterglow period. 
       
2) In order to simulate the dynamics of the microparticle cloud during the post discharge (afterglow), the parabolic confinement  was suddenly removed (corresponding to the discharge switching off) and the microparticle charges were exponentially decreased from their ``plasma-on'' values to their residual charge values with a time constant equal to 1~ms, which is comparable to the decharging time scale \cite{Couedel2006}. 

\begin{figure*}[t!]
	\centering
	\includegraphics[width=0.99\textwidth]{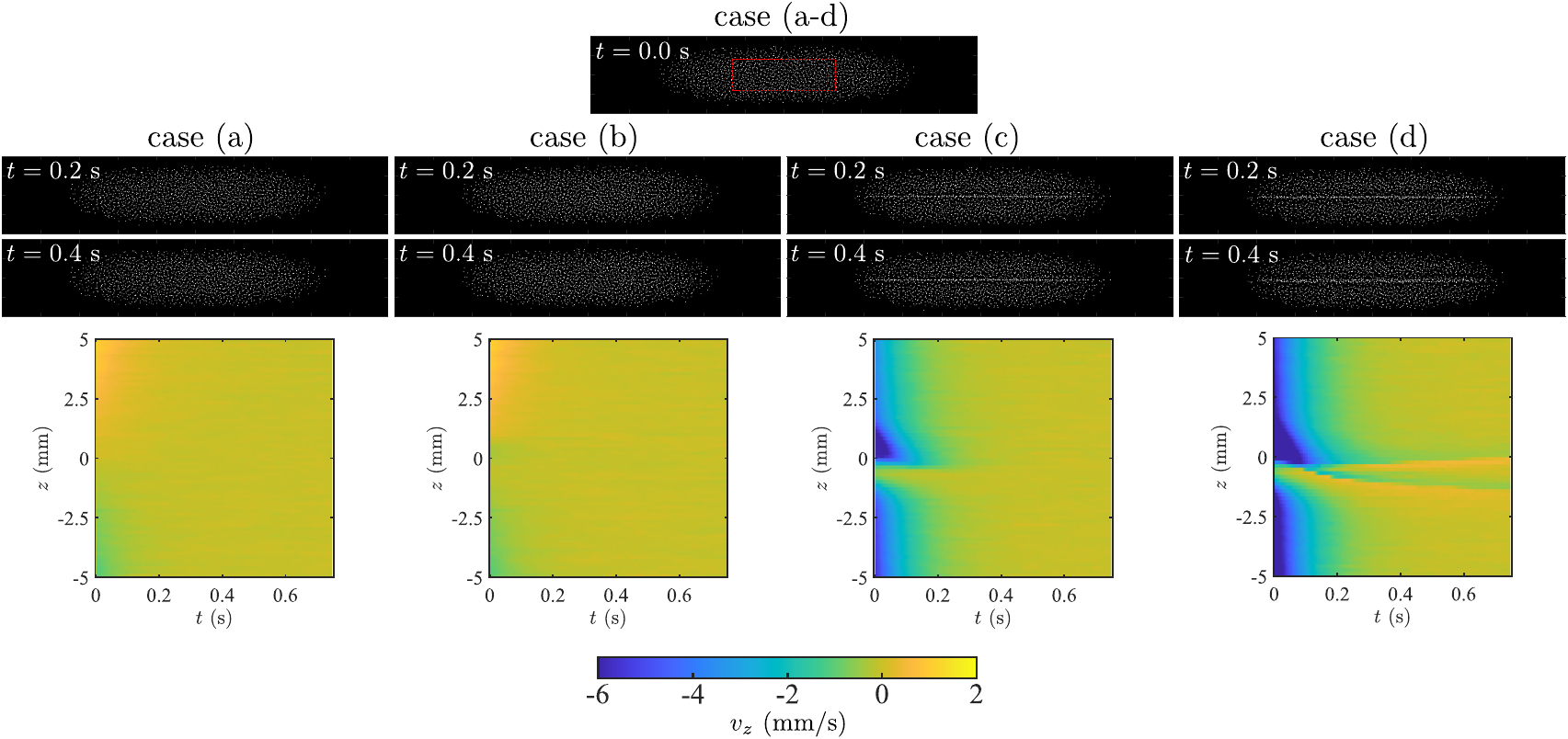} 
	\caption{Snapshots of the simulated microparticle cloud configuration at different times in the post-discharge and average vertical velocity map as a function of vertical position and time 
		for the different investigated post-discharge scenarios. In the upper snapshot at $t=0.0$~s, 
		the dashed red rectangle represents the region used to average velocity maps.}
	\label{fig:simulation_results}
\end{figure*}

For each particle, the motion was described by the following equation:
\begin{align}
        m_\rmd \frac{\rmd \vec{r}_i}{\rmd t} = \sum_{j \neq i} \vec{F}_{ij} & - m_\rmd \nu \frac{\rmd \vec{r}}{\rmd t} \nonumber \\ 
         & + Q_\mathrm{res_i}(t)\vec{E}_z (t) + L(t),
\end{align}
where  $Q_\mathrm{res_i}(t)$ is the residual charge of the $i^\mathrm{th}$ particle and $\vec{E}_z(t)$ is the external electric field due to the applied voltage on the top and bottom electrode. Note that in $\vec{F}_{ij}$ we now used $Q_\mathrm{res_i}(t)$ instead of $Q_d$.
    
Different post-discharge scenarios were investigated, as can be seen in Fig.~\ref{fig:charge_distribution}.

\begin{enumerate}[label=(\alph*)]
        \item{A homogeneous spatial distribution of microparticle charges was considered both in the ``plasma-on'' phase (i.e., at $t=0$~s) and in the afterglow (see Fig.~\ref{fig:charge_distribution}). No external electric field was imposed between the top and bottom electrodes, and therefore the microparticle residual charges were taken equal to zero. Under these conditions no particle pulse was observed during the post-discharge phase (see Fig.~\ref{fig:simulation_results}). Only a brief expansion of the dust cloud could be seen due to coulomb repulsion at the early stage of the
        post-discharge phase when the microparticle had not yet reached their residual charges. The particle motion was quickly damped due to Epstein drag. This simulation serves as baseline for the post-discharge dynamics of the microparticle cloud.}
        
        \item{The presence of a double-layer in the plasma can lead to localized spatial inhomogeneities of the dust particle charges \cite{TVM.PRE.2004} which we supposed conserved when  the discharge was switched off. This resulted in two well-defined regions of oppositely charged microparticles in the late afterglow (see Fig.~\ref{fig:charge_distribution}). In our simulation, the ``double-layer'' was arbitrarily located at the mid-plane of simulation box. The width of the ``double layer'' was arbitrarily set at 10$\lambda_\mathrm{D}$. No external field was applied in between the electrodes ($\vec{E}_z(t)=0$). In this case, the dynamics of the microparticle cloud was very similar to that of case (a), see Fig.~\ref{fig:simulation_results}.}
        
        \item{An external alternating current (AC) electric field imposed by a voltage difference between the top and bottom electrodes was added to case (b) ($\vec{E}_z(t)= 
        \vec{E}_{0_z} \cos(2 \pi f_{AC} t)$ with $\vec{E}_{0_z}=10 \hat{z}\ \mathrm{V/cm}$ and $f_{AC}=100$~Hz). In this case, in the early stage of the 
        pulse discharge, on top of the Coulomb expansion of the microparticle cloud, a motion of the highly charged particles just above the mid-plane due to the strong applied external field led to the appearance of a structure similar to the particle pulse observed in the experiments, see Fig.~\ref{fig:simulation_results}. However due to neutral drag the motion of the particles was quickly damped, and the structure did not propagate in the microparticle cloud. Moreover, the frequency of the AC was too high to trigger oscillations of the particles having a residual charge.}
        
        \item{A small external direct current (DC) field was applied on top of the AC field 
        such as $\vec{E}_z(t)= \vec{E}_{DC} + 
        \vec{E}_{0_z} \cos(2 \pi f_{AC} t)$ with $\vec{E}_{DC}=3.33 \hat{z}\ \mathrm{V/cm}$, $\vec{E}_{0_z}=10 \hat{z}\ \mathrm{V/cm}$ 
        and $f_{AC}=100$~Hz. In this case, we still observed the formation of 
        the particle pulse but, due to the electric force applied on the particles 
        having a residual charge, the pulse propagates through the cloud as observed 
        in the experiments, see Fig.~\ref{fig:simulation_results}.}
\end{enumerate}

The simulations show that only case (d) in which an additional DC field triggered the sustained motion of the microparticle pulse accurately mimicked the experiments. However, in the experiments, no DC field was applied between the electrodes, and the decay time of the field due to the self-bias voltage of the electrode disappeared on the order of a few~ns according 
to the design of PK-3 Plus matching network. Nevertheless without a source of energy, the 
motion of the particle should quickly stop (as seen in case (c)) which is clearly not 
the case in the experiments. Unless the measurements of the electrode voltage in the 
post-discharge phase were inaccurate the nature of motion of the particle pulse in the late post-discharge remains an enigma.

\section{Conclusion}
{\color{red}
To conclude, we have observed for the first time the spontaneous generation of nonlinear dust pulse structure in the plasma afterglow under microgravity condition on board the ISS. The presence of an additionally applied electric field with a low frequency oscillation at the moment when the plasma extinguished was the key to generating such phenomena. The observed nonlinear pulse structure was symmetric w.r.t the void. Similar pulse structures were observed in various missions which shows the generic nature of the phenomenon irrespective of varying experimental conditions with diverse microparticle sizes, gas pressures and types and applied voltages. 

The formation of the pulse structures could be classified depending on the relative particle flow: (1) uni-directional flow of one group of particles through a second group of stationary particles, and (2) bi-directional flow of two groups of particles through each other. The pulse propagation speed for the first case was less than for the second case. We performed both PIV and single particle identification and tracking as well as molecular dynamics simulations to understand the dynamics of the particle cloud when the plasma switched off. We attribute the counter propagation of particles to the presence of a double layer resulting in a strong spatial anisotropy of the microparticle charges. This created strong asymmetric forces on the dust particles during the early instant of the post-discharge due to the electric field created by the presence of the function generator. This hypothesis is well supported by molecular
dynamics simulation. 

The appearance of the bi-directional pulse might be due to the presence of two oppositely charged particle layers due to the impact of the double layer on the decharging process, while the emergence of the uni-directional pulse could be caused by a narrower residual
charge distribution in the post-discharge phase. However, the lasting motion of 
the particle pulse remain a mystery since at the damping rate of the experiments they would require an additional source of energy to be maintained, such as from the ions \cite{Kananovich2020a}. Further studies are therefore needed to understand the anisotropy of the microparticle charge distribution and the motion of the microparticle in the late post-discharge.

\acknowledgments

The PK-3 Plus project was funded by the space agency of the Deutsches Zentrum f\"ur Luft- und Raumfahrt eV with funds from the Federal Ministry for Economy and Technology according to a resolution of the Deutscher Bundestag under grant number 50 WM 1203.}
{\color{violet}
L. Cou\"edel acknowledges the support of the Natural Sciences and Engineering Research Council of Canada (NSERC), RGPIN-2019-04333. E. Thomas acknowledges support of NASA/JPL Grant No. JPL-RSA 1571699, U.S. Department of Energy, Grant No. DE-SC-0019176, and the National Science Foundation EPSCoR program, Cooperative Agreement No. OIA-1655280. A. M. Lipaev was supported by the Ministry of Science and Higher Education of the Russian Federation (State Assignment No. 075-00892-20-01). The authors acknowledge Roscosmos for the PK-3 Plus laboratory launch and operation on board of ISS. 

{\color{blue}We thank Victoriya Yaroshenko for careful reading of the paper, and Tanja Hagl for patiently answering our questions and generally supporting the missions. \\

We dedicate this work to our deceased colleagues Dr. V. I. Molotkov and Prof. V. E. Fortov. \\


\appendix*

{\color{red} \section{\label{sec:chargeforce}Charge and force estimation before extinction of the plasma} 
The charge (floating potential) of the particle is calculated by equating the ion flux with electron flux on its surface, $I_i = I_e$. To calculate these fluxes, typically the orbital motion limited (OML) theory is used \cite{Allen_1992,kennedy_allen_2003}. However, this approach assumes collisionless, ballistic ion and electron trajectories in the vicinity of an isolated particle without the presence of any kind of interaction potential barrier. But the collisionality index ($\zeta = \lambda_{\rm D}/\ell_i$) values in table-II shows that our experiments are performed in weakly collisional (WC) plasma regime ($\ell_i > \lambda_{\rm D}$). Here, $\lambda_D$ is the effective debye length ($\lambda_{\rm D}^{-2} = \lambda_{\rm De}^{-2} + \lambda_{\rm Di}^{-2})$ and $\ell_i$ is ion mean free path. In this regime, the OML expression for ion flux can not be applied but rather the drift motion limited ion flux can be used, 
$I_i \approx \sqrt{8\pi}a^2n_iv_{\rm Ti}[1 + z\tau + 0.1\zeta (z\tau)^2]$ where $a$ is the particle radius, $n_i$ is the ion density, $v_{\rm Ti} = \sqrt{T_i/m_i}$ is the ion thermal velocity, $z = Ze^2/aT_e$ is the normalized particle charge, $\tau = T_e/T_i$ is the electron-to-ion temperature ratio. The electron flux can be written as, $I_e \approx \sqrt{8\pi}a^2 n_e v_{Te}\exp(-z)$ where $n_e$ is the electron density, $v_{Te} = \sqrt{T_e/m_e}$ is the electron thermal velocity. Earlier investigations show that the charge value in the WC regime is significantly less than that calculated using OML theory \cite{PhysRevLett.93.085001.Ratynskaia, PhysRevE.72.016406.Khrapak,Khrapak_2012}.

To estimate the neutral drag force, we first checked the parameter regime and found that Knudsen number, $K_n (= \ell_n/a) >> 1$ and $u_d < v_{Tn}$. Here $\ell_n$ is the neutral mean-free path, $u_d$ is the particle velocity and $v_{Tn} = \sqrt{T_n/m_n}$ is the neutral thermal velocity where $m_n$ and $T_n$ are the mass and temperature of neutral gas atoms. In this parameter regime we consider the standard Epstein drag theory \cite{PhysRev.23.710.epstein},
\begin{equation}
 \label{nd} 
 F_{\rm nd} = -m_d \gamma_{ep} (u_d - u_n) 
\end{equation}
Here $\gamma_{ep} = (8\sqrt{2\pi}/3)\delta a^2 (m_n/m_d) n_n v_{Tn}$ is the Epstein frequency, where $n_n$ and $u_n$ are the density and mean velocity of neutrals respectively. Since the experiment is performed without gas flow, we consider $u_n = 0$. The numerical factor $\delta \sim 1.44$ is chosen for diffuse scattering which is consistent with complex plasma experiments \cite{liu2003,Jung2015}.   
}

 \bibliography{afterglow}%

\begin{thebibliography}{71}%
\makeatletter
\providecommand \@ifxundefined [1]{%
 \@ifx{#1\undefined}
}%
\providecommand \@ifnum [1]{%
 \ifnum #1\expandafter \@firstoftwo
 \else \expandafter \@secondoftwo
 \fi
}%
\providecommand \@ifx [1]{%
 \ifx #1\expandafter \@firstoftwo
 \else \expandafter \@secondoftwo
 \fi
}%
\providecommand \natexlab [1]{#1}%
\providecommand \enquote  [1]{``#1''}%
\providecommand \bibnamefont  [1]{#1}%
\providecommand \bibfnamefont [1]{#1}%
\providecommand \citenamefont [1]{#1}%
\providecommand \href@noop [0]{\@secondoftwo}%
\providecommand \href [0]{\begingroup \@sanitize@url \@href}%
\providecommand \@href[1]{\@@startlink{#1}\@@href}%
\providecommand \@@href[1]{\endgroup#1\@@endlink}%
\providecommand \@sanitize@url [0]{\catcode `\\12\catcode `\$12\catcode
  `\&12\catcode `\#12\catcode `\^12\catcode `\_12\catcode `\%12\relax}%
\providecommand \@@startlink[1]{}%
\providecommand \@@endlink[0]{}%
\providecommand \url  [0]{\begingroup\@sanitize@url \@url }%
\providecommand \@url [1]{\endgroup\@href {#1}{\urlprefix }}%
\providecommand \urlprefix  [0]{URL }%
\providecommand \Eprint [0]{\href }%
\providecommand \doibase [0]{https://doi.org/}%
\providecommand \selectlanguage [0]{\@gobble}%
\providecommand \bibinfo  [0]{\@secondoftwo}%
\providecommand \bibfield  [0]{\@secondoftwo}%
\providecommand \translation [1]{[#1]}%
\providecommand \BibitemOpen [0]{}%
\providecommand \bibitemStop [0]{}%
\providecommand \bibitemNoStop [0]{.\EOS\space}%
\providecommand \EOS [0]{\spacefactor3000\relax}%
\providecommand \BibitemShut  [1]{\csname bibitem#1\endcsname}%
\let\auto@bib@innerbib\@empty
\bibitem [{\citenamefont {Morfill}\ and\ \citenamefont
  {Ivlev}(2009)}]{Morfill2009}%
  \BibitemOpen
  \bibfield  {author} {\bibinfo {author} {\bibfnamefont {G.~E.}\ \bibnamefont
  {Morfill}}\ and\ \bibinfo {author} {\bibfnamefont {A.~V.}\ \bibnamefont
  {Ivlev}},\ }\bibfield  {title} {\bibinfo {title} {Complex plasmas: An
  interdisciplinary research field},\ }\href
  {https://doi.org/10.1103/RevModPhys.81.1353} {\bibfield  {journal} {\bibinfo
  {journal} {Rev. Mod. Phys.}\ }\textbf {\bibinfo {volume} {81}},\ \bibinfo
  {pages} {1353} (\bibinfo {year} {2009})}\BibitemShut {NoStop}%
\bibitem [{\citenamefont {Thomas}\ \emph {et~al.}(1994)\citenamefont {Thomas},
  \citenamefont {Morfill}, \citenamefont {Demmel}, \citenamefont {Goree},
  \citenamefont {Feuerbacher},\ and\ \citenamefont
  {M\"{o}hlmann}}]{ThomasPRL1994}%
  \BibitemOpen
  \bibfield  {author} {\bibinfo {author} {\bibfnamefont {H.}~\bibnamefont
  {Thomas}}, \bibinfo {author} {\bibfnamefont {G.~E.}\ \bibnamefont {Morfill}},
  \bibinfo {author} {\bibfnamefont {V.}~\bibnamefont {Demmel}}, \bibinfo
  {author} {\bibfnamefont {J.}~\bibnamefont {Goree}}, \bibinfo {author}
  {\bibfnamefont {B.}~\bibnamefont {Feuerbacher}},\ and\ \bibinfo {author}
  {\bibfnamefont {D.}~\bibnamefont {M\"{o}hlmann}},\ }\bibfield  {title}
  {\bibinfo {title} {Plasma crystal: Coulomb crystallization in a dusty
  plasma},\ }\href {https://doi.org/10.1103/PhysRevLett.73.652} {\bibfield
  {journal} {\bibinfo  {journal} {Phys. Rev. Lett.}\ }\textbf {\bibinfo
  {volume} {73}},\ \bibinfo {pages} {652} (\bibinfo {year} {1994})}\BibitemShut
  {NoStop}%
\bibitem [{\citenamefont {Chu}\ and\ \citenamefont {I}(1994)}]{Chu1994PRL}%
  \BibitemOpen
  \bibfield  {author} {\bibinfo {author} {\bibfnamefont {J.~H.}\ \bibnamefont
  {Chu}}\ and\ \bibinfo {author} {\bibfnamefont {L.}~\bibnamefont {I}},\
  }\bibfield  {title} {\bibinfo {title} {Direct observation of coulomb crystals
  and liquids in strongly coupled rf dusty plasmas},\ }\href
  {https://doi.org/10.1103/PhysRevLett.72.4009} {\bibfield  {journal} {\bibinfo
   {journal} {Phys. Rev. Lett.}\ }\textbf {\bibinfo {volume} {72}},\ \bibinfo
  {pages} {4009} (\bibinfo {year} {1994})}\BibitemShut {NoStop}%
\bibitem [{\citenamefont {Melzer}\ \emph {et~al.}(1994)\citenamefont {Melzer},
  \citenamefont {Trottenberg},\ and\ \citenamefont {Piel}}]{Melzer1994}%
  \BibitemOpen
  \bibfield  {author} {\bibinfo {author} {\bibfnamefont {A.}~\bibnamefont
  {Melzer}}, \bibinfo {author} {\bibfnamefont {T.}~\bibnamefont
  {Trottenberg}},\ and\ \bibinfo {author} {\bibfnamefont {A.}~\bibnamefont
  {Piel}},\ }\bibfield  {title} {\bibinfo {title} {Experimental determination
  of the charges on dust particles forming coulomb lattices},\ }\href
  {https://doi.org/10.1016/0375-9601(94)90144-9} {\bibfield  {journal}
  {\bibinfo  {journal} {Phys. Lett. A}\ }\textbf {\bibinfo {volume} {191}},\
  \bibinfo {pages} {301} (\bibinfo {year} {1994})}\BibitemShut {NoStop}%
\bibitem [{\citenamefont {Hayashi}\ and\ \citenamefont
  {Tachibana}(1994)}]{Hayashi_1994}%
  \BibitemOpen
  \bibfield  {author} {\bibinfo {author} {\bibfnamefont {Y.}~\bibnamefont
  {Hayashi}}\ and\ \bibinfo {author} {\bibfnamefont {K.}~\bibnamefont
  {Tachibana}},\ }\bibfield  {title} {\bibinfo {title} {Observation of
  coulomb-crystal formation from carbon particles grown in a methane plasma},\
  }\href {https://doi.org/10.1143/jjap.33.l804} {\bibfield  {journal} {\bibinfo
   {journal} {Japanese Journal of Applied Physics}\ }\textbf {\bibinfo {volume}
  {33}},\ \bibinfo {pages} {L804} (\bibinfo {year} {1994})}\BibitemShut
  {NoStop}%
\bibitem [{\citenamefont {Thomas}\ and\ \citenamefont
  {Morfill}(1996)}]{ThomasNATURE}%
  \BibitemOpen
  \bibfield  {author} {\bibinfo {author} {\bibfnamefont {H.~M.}\ \bibnamefont
  {Thomas}}\ and\ \bibinfo {author} {\bibfnamefont {G.~E.}\ \bibnamefont
  {Morfill}},\ }\bibfield  {title} {\bibinfo {title} {Melting dynamics of a
  plasma crystal},\ }\href {https://doi.org/10.1038/379806a0} {\bibfield
  {journal} {\bibinfo  {journal} {Nature}\ }\textbf {\bibinfo {volume} {379}},\
  \bibinfo {pages} {806} (\bibinfo {year} {1996})}\BibitemShut {NoStop}%
\bibitem [{\citenamefont {Fortov}\ \emph {et~al.}(2005)\citenamefont {Fortov},
  \citenamefont {Ivlev}, \citenamefont {Khrapak}, \citenamefont {Khrapak},\
  and\ \citenamefont {Morfill}}]{Fortov2005}%
  \BibitemOpen
  \bibfield  {author} {\bibinfo {author} {\bibfnamefont {V.~E.}\ \bibnamefont
  {Fortov}}, \bibinfo {author} {\bibfnamefont {A.~V.}\ \bibnamefont {Ivlev}},
  \bibinfo {author} {\bibfnamefont {S.~A.}\ \bibnamefont {Khrapak}}, \bibinfo
  {author} {\bibfnamefont {A.~G.}\ \bibnamefont {Khrapak}},\ and\ \bibinfo
  {author} {\bibfnamefont {G.~E.}\ \bibnamefont {Morfill}},\ }\bibfield
  {title} {\bibinfo {title} {Complex (dusty) plasmas: Current status, open
  issues, perspectives},\ }\href
  {https://doi.org/10.1016/j.physrep.2005.08.007} {\bibfield  {journal}
  {\bibinfo  {journal} {Phys. Rep.}\ }\textbf {\bibinfo {volume} {421}},\
  \bibinfo {pages} {1 } (\bibinfo {year} {2005})}\BibitemShut {NoStop}%
\bibitem [{\citenamefont {Shukla}\ and\ \citenamefont
  {Mamun}(2002)}]{Shukla2002}%
  \BibitemOpen
  \bibfield  {author} {\bibinfo {author} {\bibfnamefont {P.~K.}\ \bibnamefont
  {Shukla}}\ and\ \bibinfo {author} {\bibfnamefont {A.~A.}\ \bibnamefont
  {Mamun}},\ }\href@noop {} {\emph {\bibinfo {title} {Introduction to dusty
  plasma}}}\ (\bibinfo  {publisher} {IOP Publishing},\ \bibinfo {address}
  {Bristol},\ \bibinfo {year} {2002})\BibitemShut {NoStop}%
\bibitem [{\citenamefont {Vladimirov}\ \emph {et~al.}(2005)\citenamefont
  {Vladimirov}, \citenamefont {Ostrikov},\ and\ \citenamefont
  {Samarian}}]{Vladimirov2005}%
  \BibitemOpen
  \bibfield  {author} {\bibinfo {author} {\bibfnamefont {S.}~\bibnamefont
  {Vladimirov}}, \bibinfo {author} {\bibfnamefont {K.}~\bibnamefont
  {Ostrikov}},\ and\ \bibinfo {author} {\bibfnamefont {A.}~\bibnamefont
  {Samarian}},\ }\href@noop {} {\emph {\bibinfo {title} {Physics and
  Applications of Complex Plasmas}}},\ edited by\ \bibinfo {editor}
  {\bibfnamefont {L.}~\bibnamefont {Imperial~Press}}\ (\bibinfo  {publisher}
  {Imperial Press, London},\ \bibinfo {year} {2005})\BibitemShut {NoStop}%
\bibitem [{\citenamefont {Chaudhuri}\ \emph {et~al.}(2011)\citenamefont
  {Chaudhuri}, \citenamefont {Ivlev}, \citenamefont {Khrapak}, \citenamefont
  {Thomas},\ and\ \citenamefont {Morfill}}]{Chaudhuri2010}%
  \BibitemOpen
  \bibfield  {author} {\bibinfo {author} {\bibfnamefont {M.}~\bibnamefont
  {Chaudhuri}}, \bibinfo {author} {\bibfnamefont {A.~V.}\ \bibnamefont
  {Ivlev}}, \bibinfo {author} {\bibfnamefont {S.~A.}\ \bibnamefont {Khrapak}},
  \bibinfo {author} {\bibfnamefont {H.~M.}\ \bibnamefont {Thomas}},\ and\
  \bibinfo {author} {\bibfnamefont {G.~E.}\ \bibnamefont {Morfill}},\
  }\bibfield  {title} {\bibinfo {title} {Complex plasma - the plasma state of
  soft matter},\ }\href {https://doi.org/10.1039/C0SM00813C} {\bibfield
  {journal} {\bibinfo  {journal} {Soft Matter}\ }\textbf {\bibinfo {volume}
  {7}},\ \bibinfo {pages} {1287} (\bibinfo {year} {2011})}\BibitemShut
  {NoStop}%
\bibitem [{\citenamefont {Mendis}\ and\ \citenamefont
  {Horányi}(2013)}]{Mendis2013}%
  \BibitemOpen
  \bibfield  {author} {\bibinfo {author} {\bibfnamefont {D.~A.}\ \bibnamefont
  {Mendis}}\ and\ \bibinfo {author} {\bibfnamefont {M.}~\bibnamefont
  {Horányi}},\ }\bibfield  {title} {\bibinfo {title} {Dusty plasma effects in
  comets: Expectations for rosetta},\ }\href
  {https://doi.org/10.1002/rog.20005} {\bibfield  {journal} {\bibinfo
  {journal} {Rev. Geophys.}\ }\textbf {\bibinfo {volume} {51}},\ \bibinfo
  {pages} {53} (\bibinfo {year} {2013})}\BibitemShut {NoStop}%
\bibitem [{\citenamefont {Bouchoule}(1999)}]{Bouchoule1999}%
  \BibitemOpen
  \bibfield  {author} {\bibinfo {author} {\bibfnamefont {A.}~\bibnamefont
  {Bouchoule}},\ }\href@noop {} {\emph {\bibinfo {title} {Dusty Plasmas:
  Physics, Chemistry and Technological impacts in Plasma Processing}}}\
  (\bibinfo  {publisher} {Wiley},\ \bibinfo {address} {New York},\ \bibinfo
  {year} {1999})\BibitemShut {NoStop}%
\bibitem [{\citenamefont {Winter}(2004)}]{Winter2004}%
  \BibitemOpen
  \bibfield  {author} {\bibinfo {author} {\bibfnamefont {J.}~\bibnamefont
  {Winter}},\ }\bibfield  {title} {\bibinfo {title} {Dust in fusion devices—a
  multi-faceted problem connecting high- and low-temperature plasma physics},\
  }\href {https://doi.org/10.1088/0741-3335/46/12B/047} {\bibfield  {journal}
  {\bibinfo  {journal} {Plasma Phys. Controlled Fusion}\ }\textbf {\bibinfo
  {volume} {46}},\ \bibinfo {pages} {B583} (\bibinfo {year}
  {2004})}\BibitemShut {NoStop}%
\bibitem [{\citenamefont {Fortuna-Zaleśna}\ \emph {et~al.}(2017)\citenamefont
  {Fortuna-Zaleśna}, \citenamefont {Grzonka}, \citenamefont {Rubel},
  \citenamefont {Garcia-Carrasco}, \citenamefont {Widdowson}, \citenamefont
  {Baron-Wiechec}, \citenamefont {Ciupiński},\ and\ \citenamefont
  {Contributors}}]{Fortuna-Zalesna2017}%
  \BibitemOpen
  \bibfield  {author} {\bibinfo {author} {\bibfnamefont {E.}~\bibnamefont
  {Fortuna-Zaleśna}}, \bibinfo {author} {\bibfnamefont {J.}~\bibnamefont
  {Grzonka}}, \bibinfo {author} {\bibfnamefont {M.}~\bibnamefont {Rubel}},
  \bibinfo {author} {\bibfnamefont {A.}~\bibnamefont {Garcia-Carrasco}},
  \bibinfo {author} {\bibfnamefont {A.}~\bibnamefont {Widdowson}}, \bibinfo
  {author} {\bibfnamefont {A.}~\bibnamefont {Baron-Wiechec}}, \bibinfo {author}
  {\bibfnamefont {L.}~\bibnamefont {Ciupiński}},\ and\ \bibinfo {author}
  {\bibfnamefont {J.}~\bibnamefont {Contributors}},\ }\bibfield  {title}
  {\bibinfo {title} {Studies of dust from jet with the iter-like wall:
  Composition and internal structure},\ }\href
  {https://doi.org/10.1016/j.nme.2016.11.027} {\bibfield  {journal} {\bibinfo
  {journal} {Nuclear Materials and Energy}\ }\textbf {\bibinfo {volume} {12}},\
  \bibinfo {pages} {582 } (\bibinfo {year} {2017})}\BibitemShut {NoStop}%
\bibitem [{\citenamefont {Ratynskaia}\ \emph {et~al.}(2017)\citenamefont
  {Ratynskaia}, \citenamefont {Tolias}, \citenamefont {Angeli}, \citenamefont
  {Weinzettl}, \citenamefont {Matejicek}, \citenamefont {Bykov}, \citenamefont
  {Rudakov}, \citenamefont {Vignitchouk}, \citenamefont {Thorén},
  \citenamefont {Riva}, \citenamefont {Ripamonti}, \citenamefont {Morgan},
  \citenamefont {Panek},\ and\ \citenamefont {Temmerman}}]{Ratynskaia2017}%
  \BibitemOpen
  \bibfield  {author} {\bibinfo {author} {\bibfnamefont {S.}~\bibnamefont
  {Ratynskaia}}, \bibinfo {author} {\bibfnamefont {P.}~\bibnamefont {Tolias}},
  \bibinfo {author} {\bibfnamefont {M.~D.}\ \bibnamefont {Angeli}}, \bibinfo
  {author} {\bibfnamefont {V.}~\bibnamefont {Weinzettl}}, \bibinfo {author}
  {\bibfnamefont {J.}~\bibnamefont {Matejicek}}, \bibinfo {author}
  {\bibfnamefont {I.}~\bibnamefont {Bykov}}, \bibinfo {author} {\bibfnamefont
  {D.}~\bibnamefont {Rudakov}}, \bibinfo {author} {\bibfnamefont
  {L.}~\bibnamefont {Vignitchouk}}, \bibinfo {author} {\bibfnamefont
  {E.}~\bibnamefont {Thorén}}, \bibinfo {author} {\bibfnamefont
  {G.}~\bibnamefont {Riva}}, \bibinfo {author} {\bibfnamefont {D.}~\bibnamefont
  {Ripamonti}}, \bibinfo {author} {\bibfnamefont {T.}~\bibnamefont {Morgan}},
  \bibinfo {author} {\bibfnamefont {R.}~\bibnamefont {Panek}},\ and\ \bibinfo
  {author} {\bibfnamefont {G.~D.}\ \bibnamefont {Temmerman}},\ }\bibfield
  {title} {\bibinfo {title} {Tungsten dust remobilization under steady-state
  and transient plasma conditions},\ }\href
  {https://doi.org/10.1016/j.nme.2016.10.021} {\bibfield  {journal} {\bibinfo
  {journal} {Nuclear Materials and Energy}\ }\textbf {\bibinfo {volume} {12}},\
  \bibinfo {pages} {569 } (\bibinfo {year} {2017})}\BibitemShut {NoStop}%
\bibitem [{\citenamefont {Fortov}\ \emph {et~al.}(1998)\citenamefont {Fortov},
  \citenamefont {Nefedov}, \citenamefont {Vaulina}, \citenamefont {Lipaev},
  \citenamefont {Molotkov}, \citenamefont {Samarian}, \citenamefont
  {Nikitskii}, \citenamefont {Ivanov}, \citenamefont {Savin}, \citenamefont
  {Kalmykov}, \citenamefont {Solovaev},\ and\ \citenamefont
  {Vinogradov}}]{FortovJETP1998}%
  \BibitemOpen
  \bibfield  {author} {\bibinfo {author} {\bibfnamefont {V.~E.}\ \bibnamefont
  {Fortov}}, \bibinfo {author} {\bibfnamefont {A.~P.}\ \bibnamefont {Nefedov}},
  \bibinfo {author} {\bibfnamefont {O.~S.}\ \bibnamefont {Vaulina}}, \bibinfo
  {author} {\bibfnamefont {A.~M.}\ \bibnamefont {Lipaev}}, \bibinfo {author}
  {\bibfnamefont {V.~I.}\ \bibnamefont {Molotkov}}, \bibinfo {author}
  {\bibfnamefont {A.~A.}\ \bibnamefont {Samarian}}, \bibinfo {author}
  {\bibfnamefont {V.~P.}\ \bibnamefont {Nikitskii}}, \bibinfo {author}
  {\bibfnamefont {A.~I.}\ \bibnamefont {Ivanov}}, \bibinfo {author}
  {\bibfnamefont {S.~F.}\ \bibnamefont {Savin}}, \bibinfo {author}
  {\bibfnamefont {A.~V.}\ \bibnamefont {Kalmykov}}, \bibinfo {author}
  {\bibfnamefont {A.~Y.}\ \bibnamefont {Solovaev}},\ and\ \bibinfo {author}
  {\bibfnamefont {P.~V.}\ \bibnamefont {Vinogradov}},\ }\bibfield  {title}
  {\bibinfo {title} {Dusty plasma induced by solar radiation under
  microgravitational conditions: An experiment on board the mir orbiting space
  station},\ }\href {https://doi.org/10.1134/1.558598} {\bibfield  {journal}
  {\bibinfo  {journal} {Journal of Experimental and Theoretical Physics}\
  }\textbf {\bibinfo {volume} {87}},\ \bibinfo {pages} {1087} (\bibinfo {year}
  {1998})}\BibitemShut {NoStop}%
\bibitem [{\citenamefont {Morfill}\ \emph {et~al.}(1999)\citenamefont
  {Morfill}, \citenamefont {Thomas}, \citenamefont {Konopka}, \citenamefont
  {Rothermel}, \citenamefont {Zuzic}, \citenamefont {Ivlev},\ and\
  \citenamefont {Goree}}]{Morfill1999a}%
  \BibitemOpen
  \bibfield  {author} {\bibinfo {author} {\bibfnamefont {G.~E.}\ \bibnamefont
  {Morfill}}, \bibinfo {author} {\bibfnamefont {H.~M.}\ \bibnamefont {Thomas}},
  \bibinfo {author} {\bibfnamefont {U.}~\bibnamefont {Konopka}}, \bibinfo
  {author} {\bibfnamefont {H.}~\bibnamefont {Rothermel}}, \bibinfo {author}
  {\bibfnamefont {M.}~\bibnamefont {Zuzic}}, \bibinfo {author} {\bibfnamefont
  {A.}~\bibnamefont {Ivlev}},\ and\ \bibinfo {author} {\bibfnamefont
  {J.}~\bibnamefont {Goree}},\ }\bibfield  {title} {\bibinfo {title} {Condensed
  plasmas under microgravity},\ }\href
  {https://doi.org/10.1103/PhysRevLett.83.1598} {\bibfield  {journal} {\bibinfo
   {journal} {Phys. Rev. Lett.}\ }\textbf {\bibinfo {volume} {83}},\ \bibinfo
  {pages} {1598} (\bibinfo {year} {1999})}\BibitemShut {NoStop}%
\bibitem [{\citenamefont {Nefedov}\ \emph {et~al.}(2003)\citenamefont
  {Nefedov}, \citenamefont {Morfill}, \citenamefont {Fortov}, \citenamefont
  {Thomas}, \citenamefont {Rothermel}, \citenamefont {Hagl}, \citenamefont
  {Ivlev}, \citenamefont {Zuzic}, \citenamefont {Klumov}, \citenamefont
  {Lipaev}, \citenamefont {Molotkov}, \citenamefont {Petrov}, \citenamefont
  {Gidzenko}, \citenamefont {Krikalev}, \citenamefont {Shepherd}, \citenamefont
  {Ivanov}, \citenamefont {Roth}, \citenamefont {Binnenbruck}, \citenamefont
  {Goree},\ and\ \citenamefont {Semenov}}]{Nefedov2003}%
  \BibitemOpen
  \bibfield  {author} {\bibinfo {author} {\bibfnamefont {A.~P.}\ \bibnamefont
  {Nefedov}}, \bibinfo {author} {\bibfnamefont {G.~E.}\ \bibnamefont
  {Morfill}}, \bibinfo {author} {\bibfnamefont {V.~E.}\ \bibnamefont {Fortov}},
  \bibinfo {author} {\bibfnamefont {H.~M.}\ \bibnamefont {Thomas}}, \bibinfo
  {author} {\bibfnamefont {H.}~\bibnamefont {Rothermel}}, \bibinfo {author}
  {\bibfnamefont {T.}~\bibnamefont {Hagl}}, \bibinfo {author} {\bibfnamefont
  {A.~V.}\ \bibnamefont {Ivlev}}, \bibinfo {author} {\bibfnamefont
  {M.}~\bibnamefont {Zuzic}}, \bibinfo {author} {\bibfnamefont {B.~A.}\
  \bibnamefont {Klumov}}, \bibinfo {author} {\bibfnamefont {A.~M.}\
  \bibnamefont {Lipaev}}, \bibinfo {author} {\bibfnamefont {V.~I.}\
  \bibnamefont {Molotkov}}, \bibinfo {author} {\bibfnamefont {O.~F.}\
  \bibnamefont {Petrov}}, \bibinfo {author} {\bibfnamefont {Y.~P.}\
  \bibnamefont {Gidzenko}}, \bibinfo {author} {\bibfnamefont {S.~K.}\
  \bibnamefont {Krikalev}}, \bibinfo {author} {\bibfnamefont {W.}~\bibnamefont
  {Shepherd}}, \bibinfo {author} {\bibfnamefont {A.~I.}\ \bibnamefont
  {Ivanov}}, \bibinfo {author} {\bibfnamefont {M.}~\bibnamefont {Roth}},
  \bibinfo {author} {\bibfnamefont {H.}~\bibnamefont {Binnenbruck}}, \bibinfo
  {author} {\bibfnamefont {J.~A.}\ \bibnamefont {Goree}},\ and\ \bibinfo
  {author} {\bibfnamefont {Y.~P.}\ \bibnamefont {Semenov}},\ }\bibfield
  {title} {\bibinfo {title} {Pke-nefedov: plasma crystal experiments on the
  international space station},\ }\href
  {https://doi.org/10.1088/1367-2630/5/1/333} {\bibfield  {journal} {\bibinfo
  {journal} {New J. Phys.}\ }\textbf {\bibinfo {volume} {5}},\ \bibinfo {pages}
  {33} (\bibinfo {year} {2003})}\BibitemShut {NoStop}%
\bibitem [{\citenamefont {Klindworth}\ \emph {et~al.}(2004)\citenamefont
  {Klindworth}, \citenamefont {Piel}, \citenamefont {Melzer}, \citenamefont
  {Konopka}, \citenamefont {Rothermel}, \citenamefont {Tarantik},\ and\
  \citenamefont {Morfill}}]{Klindworth2004}%
  \BibitemOpen
  \bibfield  {author} {\bibinfo {author} {\bibfnamefont {M.}~\bibnamefont
  {Klindworth}}, \bibinfo {author} {\bibfnamefont {A.}~\bibnamefont {Piel}},
  \bibinfo {author} {\bibfnamefont {A.}~\bibnamefont {Melzer}}, \bibinfo
  {author} {\bibfnamefont {U.}~\bibnamefont {Konopka}}, \bibinfo {author}
  {\bibfnamefont {H.}~\bibnamefont {Rothermel}}, \bibinfo {author}
  {\bibfnamefont {K.}~\bibnamefont {Tarantik}},\ and\ \bibinfo {author}
  {\bibfnamefont {G.~E.}\ \bibnamefont {Morfill}},\ }\bibfield  {title}
  {\bibinfo {title} {Dust-free regions around langmuir probes in complex
  plasmas under microgravity},\ }\href
  {https://doi.org/10.1103/PhysRevLett.93.195002} {\bibfield  {journal}
  {\bibinfo  {journal} {Phys. Rev. Lett.}\ }\textbf {\bibinfo {volume} {93}},\
  \bibinfo {pages} {195002} (\bibinfo {year} {2004})}\BibitemShut {NoStop}%
\bibitem [{\citenamefont {Thomas}\ \emph {et~al.}(2008)\citenamefont {Thomas},
  \citenamefont {Morfill}, \citenamefont {Fortov}, \citenamefont {Ivlev},
  \citenamefont {Molotkov}, \citenamefont {Lipaev}, \citenamefont {Hagl},
  \citenamefont {Rothermel}, \citenamefont {Khrapak}, \citenamefont
  {S\"{u}tterlin}, \citenamefont {Rubin-Zuzic}, \citenamefont {Petrov},
  \citenamefont {Tokarev},\ and\ \citenamefont {Krikalev}}]{ThomasNJP}%
  \BibitemOpen
  \bibfield  {author} {\bibinfo {author} {\bibfnamefont {H.~M.}\ \bibnamefont
  {Thomas}}, \bibinfo {author} {\bibfnamefont {G.~E.}\ \bibnamefont {Morfill}},
  \bibinfo {author} {\bibfnamefont {V.~E.}\ \bibnamefont {Fortov}}, \bibinfo
  {author} {\bibfnamefont {A.~V.}\ \bibnamefont {Ivlev}}, \bibinfo {author}
  {\bibfnamefont {V.~I.}\ \bibnamefont {Molotkov}}, \bibinfo {author}
  {\bibfnamefont {A.~M.}\ \bibnamefont {Lipaev}}, \bibinfo {author}
  {\bibfnamefont {T.}~\bibnamefont {Hagl}}, \bibinfo {author} {\bibfnamefont
  {H.}~\bibnamefont {Rothermel}}, \bibinfo {author} {\bibfnamefont {S.~A.}\
  \bibnamefont {Khrapak}}, \bibinfo {author} {\bibfnamefont {R.~K.}\
  \bibnamefont {S\"{u}tterlin}}, \bibinfo {author} {\bibfnamefont
  {M.}~\bibnamefont {Rubin-Zuzic}}, \bibinfo {author} {\bibfnamefont {O.~F.}\
  \bibnamefont {Petrov}}, \bibinfo {author} {\bibfnamefont {V.~I.}\
  \bibnamefont {Tokarev}},\ and\ \bibinfo {author} {\bibfnamefont {S.~K.}\
  \bibnamefont {Krikalev}},\ }\bibfield  {title} {\bibinfo {title} {Complex
  plasma laboratory {PK-3 Plus} on the {International Space Station}},\ }\href
  {https://doi.org/10.1088/1367-2630/10/3/033036} {\bibfield  {journal}
  {\bibinfo  {journal} {New J. Phys.}\ }\textbf {\bibinfo {volume} {10}},\
  \bibinfo {pages} {033036} (\bibinfo {year} {2008})}\BibitemShut {NoStop}%
\bibitem [{\citenamefont {Himpel}\ \emph {et~al.}(2011)\citenamefont {Himpel},
  \citenamefont {Buttensch\"{o}n},\ and\ \citenamefont {Melzer}}]{Himpel2011}%
  \BibitemOpen
  \bibfield  {author} {\bibinfo {author} {\bibfnamefont {M.}~\bibnamefont
  {Himpel}}, \bibinfo {author} {\bibfnamefont {B.}~\bibnamefont
  {Buttensch\"{o}n}},\ and\ \bibinfo {author} {\bibfnamefont {A.}~\bibnamefont
  {Melzer}},\ }\bibfield  {title} {\bibinfo {title} {Three-view stereoscopy in
  dusty plasmas under microgravity: A calibration and reconstruction
  approach},\ }\href {https://doi.org/10.1063/1.3589858} {\bibfield  {journal}
  {\bibinfo  {journal} {Rev. Sci. Instr.}\ }\textbf {\bibinfo {volume} {82}},\
  \bibinfo {pages} {053706} (\bibinfo {year} {2011})}\BibitemShut {NoStop}%
\bibitem [{\citenamefont {Takahashi}\ \emph {et~al.}(2014)\citenamefont
  {Takahashi}, \citenamefont {Tonouchi}, \citenamefont {Adachi},\ and\
  \citenamefont {Totsuji}}]{Takahashi2014}%
  \BibitemOpen
  \bibfield  {author} {\bibinfo {author} {\bibfnamefont {K.}~\bibnamefont
  {Takahashi}}, \bibinfo {author} {\bibfnamefont {M.}~\bibnamefont {Tonouchi}},
  \bibinfo {author} {\bibfnamefont {S.}~\bibnamefont {Adachi}},\ and\ \bibinfo
  {author} {\bibfnamefont {H.}~\bibnamefont {Totsuji}},\ }\bibfield  {title}
  {\bibinfo {title} {Study of cylindrical dusty plasmas in pk-4j;
  experiments},\ }\href@noop {} {\bibfield  {journal} {\bibinfo  {journal}
  {Int. J. Microgravity Sci. Appl.}\ }\textbf {\bibinfo {volume} {31}},\
  \bibinfo {pages} {18} (\bibinfo {year} {2014})}\BibitemShut {NoStop}%
\bibitem [{\citenamefont {Pustylnik}\ \emph {et~al.}(2016)\citenamefont
  {Pustylnik}, \citenamefont {Fink}, \citenamefont {Nosenko}, \citenamefont
  {Antonova}, \citenamefont {Hagl}, \citenamefont {Thomas}, \citenamefont
  {Zobnin}, \citenamefont {Lipaev}, \citenamefont {Usachev}, \citenamefont
  {Molotkov}, \citenamefont {Petrov}, \citenamefont {Fortov}, \citenamefont
  {Rau}, \citenamefont {Deysenroth}, \citenamefont {Albrecht}, \citenamefont
  {Kretschmer}, \citenamefont {Thoma}, \citenamefont {Morfill}, \citenamefont
  {Seurig}, \citenamefont {Stettner}, \citenamefont {Alyamovskaya},
  \citenamefont {Orr}, \citenamefont {Kufner}, \citenamefont {Lavrenko},
  \citenamefont {Padalka}, \citenamefont {Serova}, \citenamefont
  {Samokutyayev},\ and\ \citenamefont {Christoforetti}}]{Pustylnik2016}%
  \BibitemOpen
  \bibfield  {author} {\bibinfo {author} {\bibfnamefont {M.~Y.}\ \bibnamefont
  {Pustylnik}}, \bibinfo {author} {\bibfnamefont {M.~A.}\ \bibnamefont {Fink}},
  \bibinfo {author} {\bibfnamefont {V.}~\bibnamefont {Nosenko}}, \bibinfo
  {author} {\bibfnamefont {T.}~\bibnamefont {Antonova}}, \bibinfo {author}
  {\bibfnamefont {T.}~\bibnamefont {Hagl}}, \bibinfo {author} {\bibfnamefont
  {H.~M.}\ \bibnamefont {Thomas}}, \bibinfo {author} {\bibfnamefont {A.~V.}\
  \bibnamefont {Zobnin}}, \bibinfo {author} {\bibfnamefont {A.~M.}\
  \bibnamefont {Lipaev}}, \bibinfo {author} {\bibfnamefont {A.~D.}\
  \bibnamefont {Usachev}}, \bibinfo {author} {\bibfnamefont {V.~I.}\
  \bibnamefont {Molotkov}}, \bibinfo {author} {\bibfnamefont {O.~F.}\
  \bibnamefont {Petrov}}, \bibinfo {author} {\bibfnamefont {V.~E.}\
  \bibnamefont {Fortov}}, \bibinfo {author} {\bibfnamefont {C.}~\bibnamefont
  {Rau}}, \bibinfo {author} {\bibfnamefont {C.}~\bibnamefont {Deysenroth}},
  \bibinfo {author} {\bibfnamefont {S.}~\bibnamefont {Albrecht}}, \bibinfo
  {author} {\bibfnamefont {M.}~\bibnamefont {Kretschmer}}, \bibinfo {author}
  {\bibfnamefont {M.~H.}\ \bibnamefont {Thoma}}, \bibinfo {author}
  {\bibfnamefont {G.~E.}\ \bibnamefont {Morfill}}, \bibinfo {author}
  {\bibfnamefont {R.}~\bibnamefont {Seurig}}, \bibinfo {author} {\bibfnamefont
  {A.}~\bibnamefont {Stettner}}, \bibinfo {author} {\bibfnamefont {V.~A.}\
  \bibnamefont {Alyamovskaya}}, \bibinfo {author} {\bibfnamefont
  {A.}~\bibnamefont {Orr}}, \bibinfo {author} {\bibfnamefont {E.}~\bibnamefont
  {Kufner}}, \bibinfo {author} {\bibfnamefont {E.~G.}\ \bibnamefont
  {Lavrenko}}, \bibinfo {author} {\bibfnamefont {G.~I.}\ \bibnamefont
  {Padalka}}, \bibinfo {author} {\bibfnamefont {E.~O.}\ \bibnamefont {Serova}},
  \bibinfo {author} {\bibfnamefont {A.~M.}\ \bibnamefont {Samokutyayev}},\ and\
  \bibinfo {author} {\bibfnamefont {S.}~\bibnamefont {Christoforetti}},\
  }\bibfield  {title} {\bibinfo {title} {Plasmakristall-4: New complex (dusty)
  plasma laboratory on board the international space station},\ }\href
  {https://doi.org/10.1063/1.4962696} {\bibfield  {journal} {\bibinfo
  {journal} {Rev. Sci. Instr.}\ }\textbf {\bibinfo {volume} {87}},\ \bibinfo
  {pages} {093505} (\bibinfo {year} {2016})}\BibitemShut {NoStop}%
\bibitem [{\citenamefont {Melzer}(2019)}]{Melzer2019}%
  \BibitemOpen
  \bibfield  {author} {\bibinfo {author} {\bibfnamefont {A.}~\bibnamefont
  {Melzer}},\ }\href@noop {} {\emph {\bibinfo {title} {Physics of Dusty
  Plasmas, Lecture Notes in Physics}}},\ Vol.\ \bibinfo {volume} {962}\
  (\bibinfo  {publisher} {Springer},\ \bibinfo {year} {2019})\BibitemShut
  {NoStop}%
\bibitem [{\citenamefont {Goree}\ \emph {et~al.}(1999)\citenamefont {Goree},
  \citenamefont {Morfill}, \citenamefont {Tsytovich},\ and\ \citenamefont
  {Vladimirov}}]{Goree1999}%
  \BibitemOpen
  \bibfield  {author} {\bibinfo {author} {\bibfnamefont {J.}~\bibnamefont
  {Goree}}, \bibinfo {author} {\bibfnamefont {G.~E.}\ \bibnamefont {Morfill}},
  \bibinfo {author} {\bibfnamefont {V.~N.}\ \bibnamefont {Tsytovich}},\ and\
  \bibinfo {author} {\bibfnamefont {S.~V.}\ \bibnamefont {Vladimirov}},\
  }\bibfield  {title} {\bibinfo {title} {Theory of dust voids in plasmas},\
  }\href {https://doi.org/10.1103/PhysRevE.59.7055} {\bibfield  {journal}
  {\bibinfo  {journal} {Phys. Rev. E}\ }\textbf {\bibinfo {volume} {59}},\
  \bibinfo {pages} {7055} (\bibinfo {year} {1999})}\BibitemShut {NoStop}%
\bibitem [{\citenamefont {Khrapak}\ \emph {et~al.}(2002)\citenamefont
  {Khrapak}, \citenamefont {Ivlev}, \citenamefont {Morfill},\ and\
  \citenamefont {Thomas}}]{Khrapak2002}%
  \BibitemOpen
  \bibfield  {author} {\bibinfo {author} {\bibfnamefont {S.~A.}\ \bibnamefont
  {Khrapak}}, \bibinfo {author} {\bibfnamefont {A.~V.}\ \bibnamefont {Ivlev}},
  \bibinfo {author} {\bibfnamefont {G.~E.}\ \bibnamefont {Morfill}},\ and\
  \bibinfo {author} {\bibfnamefont {H.~M.}\ \bibnamefont {Thomas}},\ }\bibfield
   {title} {\bibinfo {title} {Ion drag force in complex plasmas},\ }\href@noop
  {} {\bibfield  {journal} {\bibinfo  {journal} {Phys. Rev. E}\ }\textbf
  {\bibinfo {volume} {66}},\ \bibinfo {pages} {046414} (\bibinfo {year}
  {2002})}\BibitemShut {NoStop}%
\bibitem [{\citenamefont {Kretschmer}\ \emph {et~al.}(2005)\citenamefont
  {Kretschmer}, \citenamefont {Khrapak}, \citenamefont {Zhdanov}, \citenamefont
  {Thomas}, \citenamefont {Morfill}, \citenamefont {Fortov}, \citenamefont
  {Lipaev}, \citenamefont {Molotkov}, \citenamefont {Ivanov},\ and\
  \citenamefont {Turin}}]{Kretschmer2005}%
  \BibitemOpen
  \bibfield  {author} {\bibinfo {author} {\bibfnamefont {M.}~\bibnamefont
  {Kretschmer}}, \bibinfo {author} {\bibfnamefont {S.~A.}\ \bibnamefont
  {Khrapak}}, \bibinfo {author} {\bibfnamefont {S.~K.}\ \bibnamefont
  {Zhdanov}}, \bibinfo {author} {\bibfnamefont {H.~M.}\ \bibnamefont {Thomas}},
  \bibinfo {author} {\bibfnamefont {G.~E.}\ \bibnamefont {Morfill}}, \bibinfo
  {author} {\bibfnamefont {V.~E.}\ \bibnamefont {Fortov}}, \bibinfo {author}
  {\bibfnamefont {A.~M.}\ \bibnamefont {Lipaev}}, \bibinfo {author}
  {\bibfnamefont {V.~I.}\ \bibnamefont {Molotkov}}, \bibinfo {author}
  {\bibfnamefont {A.}~\bibnamefont {Ivanov}},\ and\ \bibinfo {author}
  {\bibfnamefont {M.}~\bibnamefont {Turin}},\ }\bibfield  {title} {\bibinfo
  {title} {Force field inside the void in complex plasmas under microgravity
  conditions},\ }\bibfield  {journal} {\bibinfo  {journal} {Phys. Rev. E}\
  }\textbf {\bibinfo {volume} {71}},\ \href
  {https://doi.org/10.1103/PhysRevE.71.056401} {10.1103/PhysRevE.71.056401}
  (\bibinfo {year} {2005})\BibitemShut {NoStop}%
\bibitem [{\citenamefont {Lipaev}\ \emph {et~al.}(2007)\citenamefont {Lipaev},
  \citenamefont {Khrapak}, \citenamefont {Molotkov}, \citenamefont {Morfill},
  \citenamefont {Fortov}, \citenamefont {Ivlev}, \citenamefont {Thomas},
  \citenamefont {Khrapak}, \citenamefont {Naumkin}, \citenamefont {Ivanov},
  \citenamefont {Tretschev},\ and\ \citenamefont {Padalka}}]{Lipaev2007}%
  \BibitemOpen
  \bibfield  {author} {\bibinfo {author} {\bibfnamefont {A.~M.}\ \bibnamefont
  {Lipaev}}, \bibinfo {author} {\bibfnamefont {S.~A.}\ \bibnamefont {Khrapak}},
  \bibinfo {author} {\bibfnamefont {V.~I.}\ \bibnamefont {Molotkov}}, \bibinfo
  {author} {\bibfnamefont {G.~E.}\ \bibnamefont {Morfill}}, \bibinfo {author}
  {\bibfnamefont {V.~E.}\ \bibnamefont {Fortov}}, \bibinfo {author}
  {\bibfnamefont {A.}~\bibnamefont {Ivlev}}, \bibinfo {author} {\bibfnamefont
  {H.~M.}\ \bibnamefont {Thomas}}, \bibinfo {author} {\bibfnamefont {A.~G.}\
  \bibnamefont {Khrapak}}, \bibinfo {author} {\bibfnamefont {V.~N.}\
  \bibnamefont {Naumkin}}, \bibinfo {author} {\bibfnamefont {A.~I.}\
  \bibnamefont {Ivanov}}, \bibinfo {author} {\bibfnamefont {S.~E.}\
  \bibnamefont {Tretschev}},\ and\ \bibinfo {author} {\bibfnamefont {G.~I.}\
  \bibnamefont {Padalka}},\ }\bibfield  {title} {\bibinfo {title} {Void closure
  in complex plasmas under microgravity conditions},\ }\href
  {https://doi.org/10.1103/PhysRevLett.98.265006} {\bibfield  {journal}
  {\bibinfo  {journal} {Phys. Rev. Lett.}\ }\textbf {\bibinfo {volume} {98}},\
  \bibinfo {pages} {265006} (\bibinfo {year} {2007})}\BibitemShut {NoStop}%
\bibitem [{\citenamefont {Samsonov}\ \emph {et~al.}(2003)\citenamefont
  {Samsonov}, \citenamefont {Morfill}, \citenamefont {Thomas}, \citenamefont
  {Hagl}, \citenamefont {Rothermel}, \citenamefont {Fortov}, \citenamefont
  {Lipaev}, \citenamefont {Molotkov}, \citenamefont {Nefedov}, \citenamefont
  {Petrov}, \citenamefont {Ivanov},\ and\ \citenamefont
  {Krikalev}}]{Samsonov2003}%
  \BibitemOpen
  \bibfield  {author} {\bibinfo {author} {\bibfnamefont {D.}~\bibnamefont
  {Samsonov}}, \bibinfo {author} {\bibfnamefont {G.}~\bibnamefont {Morfill}},
  \bibinfo {author} {\bibfnamefont {H.}~\bibnamefont {Thomas}}, \bibinfo
  {author} {\bibfnamefont {T.}~\bibnamefont {Hagl}}, \bibinfo {author}
  {\bibfnamefont {H.}~\bibnamefont {Rothermel}}, \bibinfo {author}
  {\bibfnamefont {V.}~\bibnamefont {Fortov}}, \bibinfo {author} {\bibfnamefont
  {A.}~\bibnamefont {Lipaev}}, \bibinfo {author} {\bibfnamefont
  {V.}~\bibnamefont {Molotkov}}, \bibinfo {author} {\bibfnamefont
  {A.}~\bibnamefont {Nefedov}}, \bibinfo {author} {\bibfnamefont
  {O.}~\bibnamefont {Petrov}}, \bibinfo {author} {\bibfnamefont
  {A.}~\bibnamefont {Ivanov}},\ and\ \bibinfo {author} {\bibfnamefont
  {S.}~\bibnamefont {Krikalev}},\ }\bibfield  {title} {\bibinfo {title}
  {Kinetic measurements of shock wave propagation in a three-dimensional
  complex (dusty) plasma},\ }\bibfield  {journal} {\bibinfo  {journal}
  {Physical Review E}\ }\textbf {\bibinfo {volume} {67}},\ \href
  {https://doi.org/10.1103/physreve.67.036404} {10.1103/physreve.67.036404}
  (\bibinfo {year} {2003})\BibitemShut {NoStop}%
\bibitem [{\citenamefont {Ivlev}\ \emph {et~al.}(2003)\citenamefont {Ivlev},
  \citenamefont {Kretschmer}, \citenamefont {Zuzic}, \citenamefont {Morfill},
  \citenamefont {Rothermel}, \citenamefont {Thomas}, \citenamefont {Fortov},
  \citenamefont {Molotkov}, \citenamefont {Nefedov}, \citenamefont {Lipaev},
  \citenamefont {Petrov}, \citenamefont {Baturin}, \citenamefont {Ivanov},\
  and\ \citenamefont {Goree}}]{Ivlev2003}%
  \BibitemOpen
  \bibfield  {author} {\bibinfo {author} {\bibfnamefont {A.~V.}\ \bibnamefont
  {Ivlev}}, \bibinfo {author} {\bibfnamefont {M.}~\bibnamefont {Kretschmer}},
  \bibinfo {author} {\bibfnamefont {M.}~\bibnamefont {Zuzic}}, \bibinfo
  {author} {\bibfnamefont {G.~E.}\ \bibnamefont {Morfill}}, \bibinfo {author}
  {\bibfnamefont {H.}~\bibnamefont {Rothermel}}, \bibinfo {author}
  {\bibfnamefont {H.~M.}\ \bibnamefont {Thomas}}, \bibinfo {author}
  {\bibfnamefont {V.~E.}\ \bibnamefont {Fortov}}, \bibinfo {author}
  {\bibfnamefont {V.~I.}\ \bibnamefont {Molotkov}}, \bibinfo {author}
  {\bibfnamefont {A.~P.}\ \bibnamefont {Nefedov}}, \bibinfo {author}
  {\bibfnamefont {A.~M.}\ \bibnamefont {Lipaev}}, \bibinfo {author}
  {\bibfnamefont {O.~F.}\ \bibnamefont {Petrov}}, \bibinfo {author}
  {\bibfnamefont {Y.~M.}\ \bibnamefont {Baturin}}, \bibinfo {author}
  {\bibfnamefont {A.~I.}\ \bibnamefont {Ivanov}},\ and\ \bibinfo {author}
  {\bibfnamefont {J.}~\bibnamefont {Goree}},\ }\bibfield  {title} {\bibinfo
  {title} {Decharging of complex plasmas: First kinetic observations},\ }\href
  {https://doi.org/10.1103/PhysRevLett.90.055003} {\bibfield  {journal}
  {\bibinfo  {journal} {Phys. Rev. Lett.}\ }\textbf {\bibinfo {volume} {90}},\
  \bibinfo {pages} {055003} (\bibinfo {year} {2003})}\BibitemShut {NoStop}%
\bibitem [{\citenamefont {Cou\"{e}del}\ \emph {et~al.}(2006)\citenamefont
  {Cou\"{e}del}, \citenamefont {Mikikian}, \citenamefont {Boufendi},\ and\
  \citenamefont {Samarian}}]{Couedel2006}%
  \BibitemOpen
  \bibfield  {author} {\bibinfo {author} {\bibfnamefont {L.}~\bibnamefont
  {Cou\"{e}del}}, \bibinfo {author} {\bibfnamefont {M.}~\bibnamefont
  {Mikikian}}, \bibinfo {author} {\bibfnamefont {L.}~\bibnamefont {Boufendi}},\
  and\ \bibinfo {author} {\bibfnamefont {A.~A.}\ \bibnamefont {Samarian}},\
  }\bibfield  {title} {\bibinfo {title} {Residual dust charges in discharge
  afterglow},\ }\href {https://doi.org/10.1103/PhysRevE.74.026403} {\bibfield
  {journal} {\bibinfo  {journal} {Phys. Rev. E}\ }\textbf {\bibinfo {volume}
  {74}},\ \bibinfo {pages} {026403} (\bibinfo {year} {2006})}\BibitemShut
  {NoStop}%
\bibitem [{\citenamefont {Cou\"{e}del}\ \emph
  {et~al.}(2008{\natexlab{a}})\citenamefont {Cou\"{e}del}, \citenamefont
  {Samarian}, \citenamefont {Mikikian},\ and\ \citenamefont
  {Boufendi}}]{Couedel2008}%
  \BibitemOpen
  \bibfield  {author} {\bibinfo {author} {\bibfnamefont {L.}~\bibnamefont
  {Cou\"{e}del}}, \bibinfo {author} {\bibfnamefont {A.~A.}\ \bibnamefont
  {Samarian}}, \bibinfo {author} {\bibfnamefont {M.}~\bibnamefont {Mikikian}},\
  and\ \bibinfo {author} {\bibfnamefont {L.}~\bibnamefont {Boufendi}},\
  }\bibfield  {title} {\bibinfo {title} {Influence of the ambipolar-to-free
  diffusion transition on dust particle charge in a complex plasma afterglow},\
  }\href {https://doi.org/10.1063/1.2938387} {\bibfield  {journal} {\bibinfo
  {journal} {Phys. Plasmas}\ }\textbf {\bibinfo {volume} {15}},\ \bibinfo
  {pages} {063705} (\bibinfo {year} {2008}{\natexlab{a}})}\BibitemShut
  {NoStop}%
\bibitem [{\citenamefont {Cou\"{e}del}\ \emph
  {et~al.}(2008{\natexlab{b}})\citenamefont {Cou\"{e}del}, \citenamefont
  {Samarian}, \citenamefont {Mikikian},\ and\ \citenamefont
  {Boufendi}}]{Couedel2008b}%
  \BibitemOpen
  \bibfield  {author} {\bibinfo {author} {\bibfnamefont {L.}~\bibnamefont
  {Cou\"{e}del}}, \bibinfo {author} {\bibfnamefont {A.~A.}\ \bibnamefont
  {Samarian}}, \bibinfo {author} {\bibfnamefont {M.}~\bibnamefont {Mikikian}},\
  and\ \bibinfo {author} {\bibfnamefont {L.}~\bibnamefont {Boufendi}},\
  }\bibfield  {title} {\bibinfo {title} {Dust-cloud dynamics in a complex
  plasma afterglow},\ }\href {https://doi.org/10.1109/TPS.2008.920220}
  {\bibfield  {journal} {\bibinfo  {journal} {IEEE Ttrans. Plasma Sci.}\
  }\textbf {\bibinfo {volume} {36}},\ \bibinfo {pages} {1014} (\bibinfo {year}
  {2008}{\natexlab{b}})}\BibitemShut {NoStop}%
\bibitem [{\citenamefont {Cou\"{e}del}\ \emph {et~al.}(2009)\citenamefont
  {Cou\"{e}del}, \citenamefont {Mezeghrane}, \citenamefont {Sa03ian},
  \citenamefont {Mikikian}, \citenamefont {Tessier}, \citenamefont {Cavarroc},\
  and\ \citenamefont {Boufendi}}]{Couedel2009a}%
  \BibitemOpen
  \bibfield  {author} {\bibinfo {author} {\bibfnamefont {L.}~\bibnamefont
  {Cou\"{e}del}}, \bibinfo {author} {\bibfnamefont {A.}~\bibnamefont
  {Mezeghrane}}, \bibinfo {author} {\bibfnamefont {A.}~\bibnamefont {Sa03ian}},
  \bibinfo {author} {\bibfnamefont {M.}~\bibnamefont {Mikikian}}, \bibinfo
  {author} {\bibfnamefont {Y.}~\bibnamefont {Tessier}}, \bibinfo {author}
  {\bibfnamefont {M.}~\bibnamefont {Cavarroc}},\ and\ \bibinfo {author}
  {\bibfnamefont {L.}~\bibnamefont {Boufendi}},\ }\bibfield  {title} {\bibinfo
  {title} {Complex plasma afterglow},\ }\href
  {https://doi.org/10.1002/ctpp.200910025} {\bibfield  {journal} {\bibinfo
  {journal} {Contrib. Plasma Phys.}\ }\textbf {\bibinfo {volume} {49}},\
  \bibinfo {pages} {235} (\bibinfo {year} {2009})}\BibitemShut {NoStop}%
\bibitem [{\citenamefont {Layden}\ \emph {et~al.}(2011)\citenamefont {Layden},
  \citenamefont {Couedel}, \citenamefont {Samarian},\ and\ \citenamefont
  {L.Boufendi}}]{Layden2011a}%
  \BibitemOpen
  \bibfield  {author} {\bibinfo {author} {\bibfnamefont {B.}~\bibnamefont
  {Layden}}, \bibinfo {author} {\bibfnamefont {L.}~\bibnamefont {Couedel}},
  \bibinfo {author} {\bibfnamefont {A.~A.}\ \bibnamefont {Samarian}},\ and\
  \bibinfo {author} {\bibnamefont {L.Boufendi}},\ }\bibfield  {title} {\bibinfo
  {title} {Residual dust charges in a complex plasma afterglow},\ }\href@noop
  {} {\bibfield  {journal} {\bibinfo  {journal} {IEEE Trans. Plas. Sci.}\
  }\textbf {\bibinfo {volume} {39}},\ \bibinfo {pages} {2764} (\bibinfo {year}
  {2011})}\BibitemShut {NoStop}%
\bibitem [{\citenamefont {Denysenko}\ \emph {et~al.}(2011)\citenamefont
  {Denysenko}, \citenamefont {Stefanovi{\'{c}}}, \citenamefont {Sikimi{\'{c}}},
  \citenamefont {Winter}, \citenamefont {Azarenkov},\ and\ \citenamefont
  {Sadeghi}}]{Denysenko2011}%
  \BibitemOpen
  \bibfield  {author} {\bibinfo {author} {\bibfnamefont {I.}~\bibnamefont
  {Denysenko}}, \bibinfo {author} {\bibfnamefont {I.}~\bibnamefont
  {Stefanovi{\'{c}}}}, \bibinfo {author} {\bibfnamefont {B.}~\bibnamefont
  {Sikimi{\'{c}}}}, \bibinfo {author} {\bibfnamefont {J.}~\bibnamefont
  {Winter}}, \bibinfo {author} {\bibfnamefont {N.~A.}\ \bibnamefont
  {Azarenkov}},\ and\ \bibinfo {author} {\bibfnamefont {N.}~\bibnamefont
  {Sadeghi}},\ }\bibfield  {title} {\bibinfo {title} {A global model for the
  afterglow of pure argon and of argon with negatively charged dust
  particles},\ }\href {https://doi.org/10.1088/0022-3727/44/20/205204}
  {\bibfield  {journal} {\bibinfo  {journal} {Journal of Physics D: Applied
  Physics}\ }\textbf {\bibinfo {volume} {44}},\ \bibinfo {pages} {205204}
  (\bibinfo {year} {2011})}\BibitemShut {NoStop}%
\bibitem [{\citenamefont {Denysenko}\ \emph {et~al.}(2013)\citenamefont
  {Denysenko}, \citenamefont {Stefanovi\`{c}}, \citenamefont {Sikimi\`{c}},
  \citenamefont {Winter},\ and\ \citenamefont {Azarenkov}}]{Denysenko2013}%
  \BibitemOpen
  \bibfield  {author} {\bibinfo {author} {\bibfnamefont {I.~B.}\ \bibnamefont
  {Denysenko}}, \bibinfo {author} {\bibfnamefont {I.}~\bibnamefont
  {Stefanovi\`{c}}}, \bibinfo {author} {\bibfnamefont {B.}~\bibnamefont
  {Sikimi\`{c}}}, \bibinfo {author} {\bibfnamefont {J.}~\bibnamefont
  {Winter}},\ and\ \bibinfo {author} {\bibfnamefont {N.~A.}\ \bibnamefont
  {Azarenkov}},\ }\bibfield  {title} {\bibinfo {title} {Discharging of dust
  particles in the afterglow of plasma with large dust density},\ }\href
  {https://doi.org/10.1103/PhysRevE.88.023104} {\bibfield  {journal} {\bibinfo
  {journal} {Phys. Rev. E}\ }\textbf {\bibinfo {volume} {88}},\ \bibinfo
  {pages} {023104} (\bibinfo {year} {2013})}\BibitemShut {NoStop}%
\bibitem [{\citenamefont {W\"{o}rner}\ \emph {et~al.}(2013)\citenamefont
  {W\"{o}rner}, \citenamefont {Ivlev}, \citenamefont {Cou\"{e}del},
  \citenamefont {Huber}, \citenamefont {Schwabe}, \citenamefont {Hagl},
  \citenamefont {Mikikian}, \citenamefont {Boufendi}, \citenamefont
  {Skvortsov}, \citenamefont {Lipaev}, \citenamefont {Molotkov}, \citenamefont
  {Petrov}, \citenamefont {Fortov}, \citenamefont {Thomas},\ and\ \citenamefont
  {Morfill}}]{Woerner2013}%
  \BibitemOpen
  \bibfield  {author} {\bibinfo {author} {\bibfnamefont {L.}~\bibnamefont
  {W\"{o}rner}}, \bibinfo {author} {\bibfnamefont {A.~V.}\ \bibnamefont
  {Ivlev}}, \bibinfo {author} {\bibfnamefont {L.}~\bibnamefont {Cou\"{e}del}},
  \bibinfo {author} {\bibfnamefont {P.}~\bibnamefont {Huber}}, \bibinfo
  {author} {\bibfnamefont {M.}~\bibnamefont {Schwabe}}, \bibinfo {author}
  {\bibfnamefont {T.}~\bibnamefont {Hagl}}, \bibinfo {author} {\bibfnamefont
  {M.}~\bibnamefont {Mikikian}}, \bibinfo {author} {\bibfnamefont
  {L.}~\bibnamefont {Boufendi}}, \bibinfo {author} {\bibfnamefont
  {A.}~\bibnamefont {Skvortsov}}, \bibinfo {author} {\bibfnamefont {A.~M.}\
  \bibnamefont {Lipaev}}, \bibinfo {author} {\bibfnamefont {V.~I.}\
  \bibnamefont {Molotkov}}, \bibinfo {author} {\bibfnamefont {O.~F.}\
  \bibnamefont {Petrov}}, \bibinfo {author} {\bibfnamefont {V.~E.}\
  \bibnamefont {Fortov}}, \bibinfo {author} {\bibfnamefont {H.~M.}\
  \bibnamefont {Thomas}},\ and\ \bibinfo {author} {\bibfnamefont {G.~E.}\
  \bibnamefont {Morfill}},\ }\bibfield  {title} {\bibinfo {title} {The effect
  of a direct current field on the microparticle charge in the plasma
  afterglow},\ }\href {https://doi.org/10.1063/1.4843855} {\bibfield  {journal}
  {\bibinfo  {journal} {Phys. Plasmas}\ }\textbf {\bibinfo {volume} {20}},\
  \bibinfo {pages} {123702} (\bibinfo {year} {2013})}\BibitemShut {NoStop}%
\bibitem [{\citenamefont {Meyer}\ and\ \citenamefont
  {Merlino}(2016)}]{Meyer2016}%
  \BibitemOpen
  \bibfield  {author} {\bibinfo {author} {\bibfnamefont {J.~K.}\ \bibnamefont
  {Meyer}}\ and\ \bibinfo {author} {\bibfnamefont {R.~L.}\ \bibnamefont
  {Merlino}},\ }\bibfield  {title} {\bibinfo {title} {Evolution of dust clouds
  in afterglow plasmas},\ }\href {https://doi.org/10.1109/TPS.2015.2504920}
  {\bibfield  {journal} {\bibinfo  {journal} {IEEE Trans. Plasma Sci.}\
  }\textbf {\bibinfo {volume} {44}},\ \bibinfo {pages} {473} (\bibinfo {year}
  {2016})}\BibitemShut {NoStop}%
\bibitem [{\citenamefont {Merlino}\ \emph {et~al.}(2016)\citenamefont
  {Merlino}, \citenamefont {Meyer}, \citenamefont {Avinash},\ and\
  \citenamefont {Sen}}]{Merlino2016}%
  \BibitemOpen
  \bibfield  {author} {\bibinfo {author} {\bibfnamefont {R.~L.}\ \bibnamefont
  {Merlino}}, \bibinfo {author} {\bibfnamefont {J.~K.}\ \bibnamefont {Meyer}},
  \bibinfo {author} {\bibfnamefont {K.}~\bibnamefont {Avinash}},\ and\ \bibinfo
  {author} {\bibfnamefont {A.}~\bibnamefont {Sen}},\ }\bibfield  {title}
  {\bibinfo {title} {Coulomb fission of a dusty plasma},\ }\bibfield  {journal}
  {\bibinfo  {journal} {Phys. Plasmas}\ }\textbf {\bibinfo {volume} {23}},\
  \href {https://doi.org/10.1063/1.4954906} {10.1063/1.4954906} (\bibinfo
  {year} {2016})\BibitemShut {NoStop}%
\bibitem [{\citenamefont {Piel}\ \emph {et~al.}(2006)\citenamefont {Piel},
  \citenamefont {Klindworth}, \citenamefont {Arp}, \citenamefont {Melzer},\
  and\ \citenamefont {Wolter}}]{Piel2006}%
  \BibitemOpen
  \bibfield  {author} {\bibinfo {author} {\bibfnamefont {A.}~\bibnamefont
  {Piel}}, \bibinfo {author} {\bibfnamefont {M.}~\bibnamefont {Klindworth}},
  \bibinfo {author} {\bibfnamefont {O.}~\bibnamefont {Arp}}, \bibinfo {author}
  {\bibfnamefont {A.}~\bibnamefont {Melzer}},\ and\ \bibinfo {author}
  {\bibfnamefont {M.}~\bibnamefont {Wolter}},\ }\bibfield  {title} {\bibinfo
  {title} {Obliquely propagating dust-density plasma waves in the presence of
  an ion beam},\ }\href {https://doi.org/10.1103/PhysRevLett.97.205009}
  {\bibfield  {journal} {\bibinfo  {journal} {Phys. Rev. Lett.}\ }\textbf
  {\bibinfo {volume} {97}},\ \bibinfo {pages} {205009} (\bibinfo {year}
  {2006})}\BibitemShut {NoStop}%
\bibitem [{\citenamefont {Schwabe}\ \emph {et~al.}(2008)\citenamefont
  {Schwabe}, \citenamefont {Zhdanov}, \citenamefont {Thomas}, \citenamefont
  {Ivlev}, \citenamefont {Rubin-Zuzic}, \citenamefont {Morfill}, \citenamefont
  {Molotkov}, \citenamefont {Lipaev}, \citenamefont {Fortov},\ and\
  \citenamefont {Reiter}}]{Schwabe2008}%
  \BibitemOpen
  \bibfield  {author} {\bibinfo {author} {\bibfnamefont {M.}~\bibnamefont
  {Schwabe}}, \bibinfo {author} {\bibfnamefont {S.~K.}\ \bibnamefont
  {Zhdanov}}, \bibinfo {author} {\bibfnamefont {H.~M.}\ \bibnamefont {Thomas}},
  \bibinfo {author} {\bibfnamefont {A.~V.}\ \bibnamefont {Ivlev}}, \bibinfo
  {author} {\bibfnamefont {M.}~\bibnamefont {Rubin-Zuzic}}, \bibinfo {author}
  {\bibfnamefont {G.~E.}\ \bibnamefont {Morfill}}, \bibinfo {author}
  {\bibfnamefont {V.~I.}\ \bibnamefont {Molotkov}}, \bibinfo {author}
  {\bibfnamefont {A.~M.}\ \bibnamefont {Lipaev}}, \bibinfo {author}
  {\bibfnamefont {V.~E.}\ \bibnamefont {Fortov}},\ and\ \bibinfo {author}
  {\bibfnamefont {T.}~\bibnamefont {Reiter}},\ }\bibfield  {title} {\bibinfo
  {title} {Nonlinear waves externally excited in a complex plasma under
  microgravity conditions},\ }\href
  {https://doi.org/10.1088/1367-2630/10/3/033037} {\bibfield  {journal}
  {\bibinfo  {journal} {New J. Phys.}\ }\textbf {\bibinfo {volume} {10}},\
  \bibinfo {pages} {033037} (\bibinfo {year} {2008})}\BibitemShut {NoStop}%
\bibitem [{\citenamefont {Khrapak}\ \emph {et~al.}(2016)\citenamefont
  {Khrapak}, \citenamefont {Molotkov}, \citenamefont {Lipaev}, \citenamefont
  {Zhukhovitskii}, \citenamefont {Naumkin}, \citenamefont {Fortov},
  \citenamefont {Petrov}, \citenamefont {Thomas}, \citenamefont {Khrapak},
  \citenamefont {Huber}, \citenamefont {Ivlev},\ and\ \citenamefont
  {Morfill}}]{Khrapak2016}%
  \BibitemOpen
  \bibfield  {author} {\bibinfo {author} {\bibfnamefont {A.~G.}\ \bibnamefont
  {Khrapak}}, \bibinfo {author} {\bibfnamefont {V.~I.}\ \bibnamefont
  {Molotkov}}, \bibinfo {author} {\bibfnamefont {A.~M.}\ \bibnamefont
  {Lipaev}}, \bibinfo {author} {\bibfnamefont {D.~I.}\ \bibnamefont
  {Zhukhovitskii}}, \bibinfo {author} {\bibfnamefont {V.~N.}\ \bibnamefont
  {Naumkin}}, \bibinfo {author} {\bibfnamefont {V.~E.}\ \bibnamefont {Fortov}},
  \bibinfo {author} {\bibfnamefont {O.~F.}\ \bibnamefont {Petrov}}, \bibinfo
  {author} {\bibfnamefont {H.~M.}\ \bibnamefont {Thomas}}, \bibinfo {author}
  {\bibfnamefont {S.~A.}\ \bibnamefont {Khrapak}}, \bibinfo {author}
  {\bibfnamefont {P.}~\bibnamefont {Huber}}, \bibinfo {author} {\bibfnamefont
  {A.}~\bibnamefont {Ivlev}},\ and\ \bibinfo {author} {\bibfnamefont
  {G.}~\bibnamefont {Morfill}},\ }\bibfield  {title} {\bibinfo {title} {Complex
  plasma research under microgravity conditions: Pk-3 plus laboratory on the
  international space station},\ }\href
  {https://doi.org/10.1002/ctpp.201500102} {\bibfield  {journal} {\bibinfo
  {journal} {Contrib. Plasma Phys.}\ }\textbf {\bibinfo {volume} {56}},\
  \bibinfo {pages} {253} (\bibinfo {year} {2016})}\BibitemShut {NoStop}%
\bibitem [{\citenamefont {Schwabe}\ \emph {et~al.}(2018)\citenamefont
  {Schwabe}, \citenamefont {Du}, \citenamefont {Huber}, \citenamefont {Lipaev},
  \citenamefont {Molotkov}, \citenamefont {Naumkin}, \citenamefont {Zhdanov},
  \citenamefont {Zhukhovitskii}, \citenamefont {Fortov},\ and\ \citenamefont
  {Thomas}}]{Schwabe2018}%
  \BibitemOpen
  \bibfield  {author} {\bibinfo {author} {\bibfnamefont {M.}~\bibnamefont
  {Schwabe}}, \bibinfo {author} {\bibfnamefont {C.-R.}\ \bibnamefont {Du}},
  \bibinfo {author} {\bibfnamefont {P.}~\bibnamefont {Huber}}, \bibinfo
  {author} {\bibfnamefont {A.~M.}\ \bibnamefont {Lipaev}}, \bibinfo {author}
  {\bibfnamefont {V.~I.}\ \bibnamefont {Molotkov}}, \bibinfo {author}
  {\bibfnamefont {V.~N.}\ \bibnamefont {Naumkin}}, \bibinfo {author}
  {\bibfnamefont {S.~K.}\ \bibnamefont {Zhdanov}}, \bibinfo {author}
  {\bibfnamefont {D.~I.}\ \bibnamefont {Zhukhovitskii}}, \bibinfo {author}
  {\bibfnamefont {V.~E.}\ \bibnamefont {Fortov}},\ and\ \bibinfo {author}
  {\bibfnamefont {H.~M.}\ \bibnamefont {Thomas}},\ }\bibfield  {title}
  {\bibinfo {title} {Latest results on complex plasmas with the {PK-3 Plus
  Laboratory} on board the {International Space Station}},\ }\href
  {https://doi.org/10.1007/s12217-018-9602-0} {\bibfield  {journal} {\bibinfo
  {journal} {Micrograv. Sci. Technol.}\ }\textbf {\bibinfo {volume} {30}},\
  \bibinfo {pages} {581} (\bibinfo {year} {2018})}\BibitemShut {NoStop}%
\bibitem [{\citenamefont {Thomas}\ \emph {et~al.}(2019)\citenamefont {Thomas},
  \citenamefont {Schwabe}, \citenamefont {Pustylnik}, \citenamefont {Knapek},
  \citenamefont {Molotkov}, \citenamefont {Lipaev}, \citenamefont {Petrov},
  \citenamefont {Fortov},\ and\ \citenamefont {Khrapak}}]{Thomas2019}%
  \BibitemOpen
  \bibfield  {author} {\bibinfo {author} {\bibfnamefont {H.~M.}\ \bibnamefont
  {Thomas}}, \bibinfo {author} {\bibfnamefont {M.}~\bibnamefont {Schwabe}},
  \bibinfo {author} {\bibfnamefont {M.}~\bibnamefont {Pustylnik}}, \bibinfo
  {author} {\bibfnamefont {C.~A.}\ \bibnamefont {Knapek}}, \bibinfo {author}
  {\bibfnamefont {V.~I.}\ \bibnamefont {Molotkov}}, \bibinfo {author}
  {\bibfnamefont {A.~M.}\ \bibnamefont {Lipaev}}, \bibinfo {author}
  {\bibfnamefont {O.~F.}\ \bibnamefont {Petrov}}, \bibinfo {author}
  {\bibfnamefont {V.~E.}\ \bibnamefont {Fortov}},\ and\ \bibinfo {author}
  {\bibfnamefont {S.}~\bibnamefont {Khrapak}},\ }\bibfield  {title} {\bibinfo
  {title} {Complex plasma research on the {International Space Station}},\
  }\href {https://doi.org/10.1088/1361-6587/aae468} {\bibfield  {journal}
  {\bibinfo  {journal} {Plasma Phys. Control. Fusion}\ }\textbf {\bibinfo
  {volume} {61}},\ \bibinfo {pages} {014004} (\bibinfo {year}
  {2019})}\BibitemShut {NoStop}%
\bibitem [{\citenamefont {Heidemann}\ \emph {et~al.}(2011)\citenamefont
  {Heidemann}, \citenamefont {Cou\"{e}del}, \citenamefont {Zhdanov},
  \citenamefont {S\"{u}tterlin}, \citenamefont {Schwabe}, \citenamefont
  {Thomas}, \citenamefont {Ivlev}, \citenamefont {Hagl}, \citenamefont
  {Morfill}, \citenamefont {Fortov}, \citenamefont {Molotkov}, \citenamefont
  {Petrov}, \citenamefont {Lipaev}, \citenamefont {Tokarev}, \citenamefont
  {Reiter},\ and\ \citenamefont {Vinogradov}}]{Heidemann2011}%
  \BibitemOpen
  \bibfield  {author} {\bibinfo {author} {\bibfnamefont {R.}~\bibnamefont
  {Heidemann}}, \bibinfo {author} {\bibfnamefont {L.}~\bibnamefont
  {Cou\"{e}del}}, \bibinfo {author} {\bibfnamefont {S.}~\bibnamefont
  {Zhdanov}}, \bibinfo {author} {\bibfnamefont {K.}~\bibnamefont
  {S\"{u}tterlin}}, \bibinfo {author} {\bibfnamefont {M.}~\bibnamefont
  {Schwabe}}, \bibinfo {author} {\bibfnamefont {H.}~\bibnamefont {Thomas}},
  \bibinfo {author} {\bibfnamefont {A.}~\bibnamefont {Ivlev}}, \bibinfo
  {author} {\bibfnamefont {T.}~\bibnamefont {Hagl}}, \bibinfo {author}
  {\bibfnamefont {G.}~\bibnamefont {Morfill}}, \bibinfo {author} {\bibfnamefont
  {V.}~\bibnamefont {Fortov}}, \bibinfo {author} {\bibfnamefont
  {V.}~\bibnamefont {Molotkov}}, \bibinfo {author} {\bibfnamefont
  {O.}~\bibnamefont {Petrov}}, \bibinfo {author} {\bibfnamefont
  {A.}~\bibnamefont {Lipaev}}, \bibinfo {author} {\bibfnamefont
  {V.}~\bibnamefont {Tokarev}}, \bibinfo {author} {\bibfnamefont
  {T.}~\bibnamefont {Reiter}},\ and\ \bibinfo {author} {\bibfnamefont
  {P.}~\bibnamefont {Vinogradov}},\ }\bibfield  {title} {\bibinfo {title}
  {Comprehensive experimental study of heartbeat oscillations observed under
  microgravity conditions in the pk-3 plus laboratory on board the
  international space station},\ }\href {https://doi.org/10.1063/1.3574905}
  {\bibfield  {journal} {\bibinfo  {journal} {Phys. Plasmas}\ }\textbf
  {\bibinfo {volume} {18}},\ \bibinfo {pages} {053701} (\bibinfo {year}
  {2011})}\BibitemShut {NoStop}%
\bibitem [{\citenamefont {Schwabe}\ \emph {et~al.}(2017)\citenamefont
  {Schwabe}, \citenamefont {Zhdanov},\ and\ \citenamefont
  {R\"ath}}]{Schwabe2017}%
  \BibitemOpen
  \bibfield  {author} {\bibinfo {author} {\bibfnamefont {M.}~\bibnamefont
  {Schwabe}}, \bibinfo {author} {\bibfnamefont {S.}~\bibnamefont {Zhdanov}},\
  and\ \bibinfo {author} {\bibfnamefont {C.}~\bibnamefont {R\"ath}},\
  }\bibfield  {title} {\bibinfo {title} {Instability onset and scaling laws of
  an autooscillating turbulent flow in a complex (dusty) plasma},\ }\href
  {https://doi.org/10.1103/PhysRevE.95.041201} {\bibfield  {journal} {\bibinfo
  {journal} {Phys. Rev. E}\ }\textbf {\bibinfo {volume} {95}},\ \bibinfo
  {pages} {041201(R)} (\bibinfo {year} {2017})}\BibitemShut {NoStop}%
\bibitem [{\citenamefont {Khrapak}\ \emph
  {et~al.}(2012{\natexlab{a}})\citenamefont {Khrapak}, \citenamefont {Klumov},
  \citenamefont {Huber}, \citenamefont {Molotkov}, \citenamefont {Lipaev},
  \citenamefont {Naumkin}, \citenamefont {Ivlev}, \citenamefont {Thomas},
  \citenamefont {Schwabe}, \citenamefont {Morfill}, \citenamefont {Petrov},
  \citenamefont {Fortov}, \citenamefont {Malentschenko},\ and\ \citenamefont
  {Volkov}}]{Khrapak2012b}%
  \BibitemOpen
  \bibfield  {author} {\bibinfo {author} {\bibfnamefont {S.~A.}\ \bibnamefont
  {Khrapak}}, \bibinfo {author} {\bibfnamefont {B.~A.}\ \bibnamefont {Klumov}},
  \bibinfo {author} {\bibfnamefont {P.}~\bibnamefont {Huber}}, \bibinfo
  {author} {\bibfnamefont {V.~I.}\ \bibnamefont {Molotkov}}, \bibinfo {author}
  {\bibfnamefont {A.~M.}\ \bibnamefont {Lipaev}}, \bibinfo {author}
  {\bibfnamefont {V.~N.}\ \bibnamefont {Naumkin}}, \bibinfo {author}
  {\bibfnamefont {A.~V.}\ \bibnamefont {Ivlev}}, \bibinfo {author}
  {\bibfnamefont {H.~M.}\ \bibnamefont {Thomas}}, \bibinfo {author}
  {\bibfnamefont {M.}~\bibnamefont {Schwabe}}, \bibinfo {author} {\bibfnamefont
  {G.~E.}\ \bibnamefont {Morfill}}, \bibinfo {author} {\bibfnamefont {O.~F.}\
  \bibnamefont {Petrov}}, \bibinfo {author} {\bibfnamefont {V.~E.}\
  \bibnamefont {Fortov}}, \bibinfo {author} {\bibfnamefont {Y.}~\bibnamefont
  {Malentschenko}},\ and\ \bibinfo {author} {\bibfnamefont {S.}~\bibnamefont
  {Volkov}},\ }\bibfield  {title} {\bibinfo {title} {Fluid-solid phase
  transitions in three-dimensional complex plasmas under microgravity
  conditions},\ }\href {https://doi.org/10.1103/PhysRevE.85.066407} {\bibfield
  {journal} {\bibinfo  {journal} {Phys. Rev. E}\ }\textbf {\bibinfo {volume}
  {85}},\ \bibinfo {pages} {066407} (\bibinfo {year}
  {2012}{\natexlab{a}})}\BibitemShut {NoStop}%
\bibitem [{\citenamefont {S\"{u}tterlin}\ \emph {et~al.}(2009)\citenamefont
  {S\"{u}tterlin}, \citenamefont {Wysocki}, \citenamefont {Ivlev},
  \citenamefont {R\"{a}th}, \citenamefont {Thomas}, \citenamefont
  {Rubin-Zuzic}, \citenamefont {Goedheer}, \citenamefont {Fortov},
  \citenamefont {Lipaev}, \citenamefont {Molotkov}, \citenamefont {Petrov},
  \citenamefont {Morfill},\ and\ \citenamefont {L\"{o}wen}}]{SutterlinPRL}%
  \BibitemOpen
  \bibfield  {author} {\bibinfo {author} {\bibfnamefont {K.~R.}\ \bibnamefont
  {S\"{u}tterlin}}, \bibinfo {author} {\bibfnamefont {A.}~\bibnamefont
  {Wysocki}}, \bibinfo {author} {\bibfnamefont {A.~V.}\ \bibnamefont {Ivlev}},
  \bibinfo {author} {\bibfnamefont {C.}~\bibnamefont {R\"{a}th}}, \bibinfo
  {author} {\bibfnamefont {H.~M.}\ \bibnamefont {Thomas}}, \bibinfo {author}
  {\bibfnamefont {M.}~\bibnamefont {Rubin-Zuzic}}, \bibinfo {author}
  {\bibfnamefont {W.~J.}\ \bibnamefont {Goedheer}}, \bibinfo {author}
  {\bibfnamefont {V.~E.}\ \bibnamefont {Fortov}}, \bibinfo {author}
  {\bibfnamefont {A.~M.}\ \bibnamefont {Lipaev}}, \bibinfo {author}
  {\bibfnamefont {V.~I.}\ \bibnamefont {Molotkov}}, \bibinfo {author}
  {\bibfnamefont {O.~F.}\ \bibnamefont {Petrov}}, \bibinfo {author}
  {\bibfnamefont {G.~E.}\ \bibnamefont {Morfill}},\ and\ \bibinfo {author}
  {\bibfnamefont {H.}~\bibnamefont {L\"{o}wen}},\ }\bibfield  {title} {\bibinfo
  {title} {Dynamics of lane formation in driven binary complex plasmas},\
  }\href {https://doi.org/10.1103/PhysRevLett.102.085003} {\bibfield  {journal}
  {\bibinfo  {journal} {Phys. Rev. Lett.}\ }\textbf {\bibinfo {volume} {102}},\
  \bibinfo {pages} {085003} (\bibinfo {year} {2009})}\BibitemShut {NoStop}%
\bibitem [{\citenamefont {Molotkov}\ \emph {et~al.}(2017)\citenamefont
  {Molotkov}, \citenamefont {Naumkin}, \citenamefont {Lipaev}, \citenamefont
  {Zhukhovitskii}, \citenamefont {Usachev}, \citenamefont {Fortov},\ and\
  \citenamefont {Thomas}}]{Molotkov2017}%
  \BibitemOpen
  \bibfield  {author} {\bibinfo {author} {\bibfnamefont {V.~I.}\ \bibnamefont
  {Molotkov}}, \bibinfo {author} {\bibfnamefont {V.~N.}\ \bibnamefont
  {Naumkin}}, \bibinfo {author} {\bibfnamefont {A.~M.}\ \bibnamefont {Lipaev}},
  \bibinfo {author} {\bibfnamefont {D.~I.}\ \bibnamefont {Zhukhovitskii}},
  \bibinfo {author} {\bibfnamefont {A.~D.}\ \bibnamefont {Usachev}}, \bibinfo
  {author} {\bibfnamefont {V.~E.}\ \bibnamefont {Fortov}},\ and\ \bibinfo
  {author} {\bibfnamefont {H.~M.}\ \bibnamefont {Thomas}},\ }\bibfield  {title}
  {\bibinfo {title} {Experiments on phase transitions in three-dimensional
  dusty plasma under microgravity conditions},\ }\href
  {https://doi.org/10.1088/1742-6596/927/1/012037} {\bibfield  {journal}
  {\bibinfo  {journal} {J. Phys.: Conf. Series}\ }\textbf {\bibinfo {volume}
  {927}},\ \bibinfo {pages} {012037} (\bibinfo {year} {2017})}\BibitemShut
  {NoStop}%
\bibitem [{\citenamefont {Naumkin}\ \emph {et~al.}(2018)\citenamefont
  {Naumkin}, \citenamefont {Lipaev}, \citenamefont {Molotkov}, \citenamefont
  {Zhukhovitskii}, \citenamefont {Usachev},\ and\ \citenamefont
  {Thomas}}]{Naumkin2018}%
  \BibitemOpen
  \bibfield  {author} {\bibinfo {author} {\bibfnamefont {V.~N.}\ \bibnamefont
  {Naumkin}}, \bibinfo {author} {\bibfnamefont {A.~M.}\ \bibnamefont {Lipaev}},
  \bibinfo {author} {\bibfnamefont {V.~I.}\ \bibnamefont {Molotkov}}, \bibinfo
  {author} {\bibfnamefont {D.~I.}\ \bibnamefont {Zhukhovitskii}}, \bibinfo
  {author} {\bibfnamefont {A.~D.}\ \bibnamefont {Usachev}},\ and\ \bibinfo
  {author} {\bibfnamefont {H.~M.}\ \bibnamefont {Thomas}},\ }\bibfield  {title}
  {\bibinfo {title} {Crystal{\textendash}liquid phase transitions in
  three-dimensional complex plasma under microgravity conditions},\ }\href
  {https://doi.org/10.1088/1742-6596/946/1/012144} {\bibfield  {journal}
  {\bibinfo  {journal} {J. Phys.: Conf. Series}\ }\textbf {\bibinfo {volume}
  {946}},\ \bibinfo {pages} {012144} (\bibinfo {year} {2018})}\BibitemShut
  {NoStop}%
\bibitem [{\citenamefont {Zhukhovitskii}\ \emph {et~al.}(2019)\citenamefont
  {Zhukhovitskii}, \citenamefont {Naumkin}, \citenamefont {Molotkov},
  \citenamefont {Lipaev},\ and\ \citenamefont {Thomas}}]{Zhukhovitskii2019}%
  \BibitemOpen
  \bibfield  {author} {\bibinfo {author} {\bibfnamefont {D.~I.}\ \bibnamefont
  {Zhukhovitskii}}, \bibinfo {author} {\bibfnamefont {V.~N.}\ \bibnamefont
  {Naumkin}}, \bibinfo {author} {\bibfnamefont {V.~I.}\ \bibnamefont
  {Molotkov}}, \bibinfo {author} {\bibfnamefont {A.~M.}\ \bibnamefont
  {Lipaev}},\ and\ \bibinfo {author} {\bibfnamefont {H.~M.}\ \bibnamefont
  {Thomas}},\ }\bibfield  {title} {\bibinfo {title} {New approach to
  measurement of the three-dimensional crystallization front propagation
  velocity in strongly coupled complex plasma},\ }\href
  {https://doi.org/10.1088/1361-6595/ab27a9} {\bibfield  {journal} {\bibinfo
  {journal} {Plasma Sources Science and Technology}\ }\textbf {\bibinfo
  {volume} {28}},\ \bibinfo {pages} {065014} (\bibinfo {year}
  {2019})}\BibitemShut {NoStop}%
\bibitem [{\citenamefont {Ivlev}\ \emph {et~al.}(2008)\citenamefont {Ivlev},
  \citenamefont {Morfill}, \citenamefont {Thomas}, \citenamefont {R\"{a}th},
  \citenamefont {Joyce}, \citenamefont {Huber}, \citenamefont {Kompaneets},
  \citenamefont {Fortov}, \citenamefont {Lipaev}, \citenamefont {Molotkov},
  \citenamefont {Reiter}, \citenamefont {Turin},\ and\ \citenamefont
  {Vinogradov}}]{Ivlev2008}%
  \BibitemOpen
  \bibfield  {author} {\bibinfo {author} {\bibfnamefont {A.~V.}\ \bibnamefont
  {Ivlev}}, \bibinfo {author} {\bibfnamefont {G.~E.}\ \bibnamefont {Morfill}},
  \bibinfo {author} {\bibfnamefont {H.~M.}\ \bibnamefont {Thomas}}, \bibinfo
  {author} {\bibfnamefont {C.}~\bibnamefont {R\"{a}th}}, \bibinfo {author}
  {\bibfnamefont {G.}~\bibnamefont {Joyce}}, \bibinfo {author} {\bibfnamefont
  {P.}~\bibnamefont {Huber}}, \bibinfo {author} {\bibfnamefont
  {R.}~\bibnamefont {Kompaneets}}, \bibinfo {author} {\bibfnamefont {V.~E.}\
  \bibnamefont {Fortov}}, \bibinfo {author} {\bibfnamefont {A.~M.}\
  \bibnamefont {Lipaev}}, \bibinfo {author} {\bibfnamefont {V.~I.}\
  \bibnamefont {Molotkov}}, \bibinfo {author} {\bibfnamefont {T.}~\bibnamefont
  {Reiter}}, \bibinfo {author} {\bibfnamefont {M.}~\bibnamefont {Turin}},\ and\
  \bibinfo {author} {\bibfnamefont {P.}~\bibnamefont {Vinogradov}},\ }\bibfield
   {title} {\bibinfo {title} {First observation of electrorheological
  plasmas},\ }\href {https://doi.org/10.1103/PhysRevLett.100.095003} {\bibfield
   {journal} {\bibinfo  {journal} {Phys. Rev. Lett.}\ }\textbf {\bibinfo
  {volume} {100}},\ \bibinfo {pages} {095003} (\bibinfo {year}
  {2008})}\BibitemShut {NoStop}%
\bibitem [{\citenamefont {Yang}\ \emph {et~al.}(2017)\citenamefont {Yang},
  \citenamefont {Schwabe}, \citenamefont {Zhdanov}, \citenamefont {Thomas},
  \citenamefont {Lipaev}, \citenamefont {Molotkov}, \citenamefont {Fortov},
  \citenamefont {Zhang},\ and\ \citenamefont {Du}}]{Yang2017}%
  \BibitemOpen
  \bibfield  {author} {\bibinfo {author} {\bibfnamefont {L.}~\bibnamefont
  {Yang}}, \bibinfo {author} {\bibfnamefont {M.}~\bibnamefont {Schwabe}},
  \bibinfo {author} {\bibfnamefont {S.}~\bibnamefont {Zhdanov}}, \bibinfo
  {author} {\bibfnamefont {H.~M.}\ \bibnamefont {Thomas}}, \bibinfo {author}
  {\bibfnamefont {A.~M.}\ \bibnamefont {Lipaev}}, \bibinfo {author}
  {\bibfnamefont {V.~I.}\ \bibnamefont {Molotkov}}, \bibinfo {author}
  {\bibfnamefont {V.~E.}\ \bibnamefont {Fortov}}, \bibinfo {author}
  {\bibfnamefont {J.}~\bibnamefont {Zhang}},\ and\ \bibinfo {author}
  {\bibfnamefont {C.-R.}\ \bibnamefont {Du}},\ }\bibfield  {title} {\bibinfo
  {title} {Density waves at the interface of a binary complex plasma},\ }\href
  {https://doi.org/10.1209/0295-5075/117/25001} {\bibfield  {journal} {\bibinfo
   {journal} {EPL}\ }\textbf {\bibinfo {volume} {117}},\ \bibinfo {pages}
  {25001} (\bibinfo {year} {2017})}\BibitemShut {NoStop}%
\bibitem [{\citenamefont {Sun}\ \emph {et~al.}(2018)\citenamefont {Sun},
  \citenamefont {Schwabe}, \citenamefont {Thomas}, \citenamefont {Lipaev},
  \citenamefont {Molotkov}, \citenamefont {Fortov}, \citenamefont {Feng},
  \citenamefont {Lin}, \citenamefont {Zhang},\ and\ \citenamefont
  {Du}}]{Sun2018}%
  \BibitemOpen
  \bibfield  {author} {\bibinfo {author} {\bibfnamefont {W.}~\bibnamefont
  {Sun}}, \bibinfo {author} {\bibfnamefont {M.}~\bibnamefont {Schwabe}},
  \bibinfo {author} {\bibfnamefont {H.~M.}\ \bibnamefont {Thomas}}, \bibinfo
  {author} {\bibfnamefont {A.~M.}\ \bibnamefont {Lipaev}}, \bibinfo {author}
  {\bibfnamefont {V.~I.}\ \bibnamefont {Molotkov}}, \bibinfo {author}
  {\bibfnamefont {V.~E.}\ \bibnamefont {Fortov}}, \bibinfo {author}
  {\bibfnamefont {Y.}~\bibnamefont {Feng}}, \bibinfo {author} {\bibfnamefont
  {Y.-F.}\ \bibnamefont {Lin}}, \bibinfo {author} {\bibfnamefont
  {J.}~\bibnamefont {Zhang}},\ and\ \bibinfo {author} {\bibfnamefont {C.-R.}\
  \bibnamefont {Du}},\ }\bibfield  {title} {\bibinfo {title} {Dissipative
  solitary wave at the interface of a binary complex plasma},\ }\href
  {https://doi.org/10.1209/0295-5075/122/55001} {\bibfield  {journal} {\bibinfo
   {journal} {EPL}\ }\textbf {\bibinfo {volume} {122}},\ \bibinfo {pages}
  {55001} (\bibinfo {year} {2018})}\BibitemShut {NoStop}%
\bibitem [{\citenamefont {Epstein}(1924)}]{PhysRev.23.710.epstein}%
  \BibitemOpen
  \bibfield  {author} {\bibinfo {author} {\bibfnamefont {P.~S.}\ \bibnamefont
  {Epstein}},\ }\bibfield  {title} {\bibinfo {title} {On the resistance
  experienced by spheres in their motion through gases},\ }\href
  {https://doi.org/10.1103/PhysRev.23.710} {\bibfield  {journal} {\bibinfo
  {journal} {Phys. Rev.}\ }\textbf {\bibinfo {volume} {23}},\ \bibinfo {pages}
  {710} (\bibinfo {year} {1924})}\BibitemShut {NoStop}%
\bibitem [{\citenamefont {Jung}\ \emph {et~al.}(2015)\citenamefont {Jung},
  \citenamefont {Greiner}, \citenamefont {Asnaz}, \citenamefont {Carstensen},\
  and\ \citenamefont {Piel}}]{Jung2015}%
  \BibitemOpen
  \bibfield  {author} {\bibinfo {author} {\bibfnamefont {H.}~\bibnamefont
  {Jung}}, \bibinfo {author} {\bibfnamefont {F.}~\bibnamefont {Greiner}},
  \bibinfo {author} {\bibfnamefont {O.~H.}\ \bibnamefont {Asnaz}}, \bibinfo
  {author} {\bibfnamefont {J.}~\bibnamefont {Carstensen}},\ and\ \bibinfo
  {author} {\bibfnamefont {A.}~\bibnamefont {Piel}},\ }\bibfield  {title}
  {\bibinfo {title} {Exploring the wake of a dust particle by a continuously
  approaching test grain},\ }\href {https://doi.org/10.1063/1.4920968}
  {\bibfield  {journal} {\bibinfo  {journal} {Phys. Plasmas}\ }\textbf
  {\bibinfo {volume} {22}},\ \bibinfo {pages} {053702} (\bibinfo {year}
  {2015})}\BibitemShut {NoStop}%
\bibitem [{\citenamefont {Schwabe}\ \emph {et~al.}(2007)\citenamefont
  {Schwabe}, \citenamefont {Rubin-Zuzic}, \citenamefont {Zhdanov},
  \citenamefont {Thomas},\ and\ \citenamefont {Morfill}}]{Schwabe2007}%
  \BibitemOpen
  \bibfield  {author} {\bibinfo {author} {\bibfnamefont {M.}~\bibnamefont
  {Schwabe}}, \bibinfo {author} {\bibfnamefont {M.}~\bibnamefont
  {Rubin-Zuzic}}, \bibinfo {author} {\bibfnamefont {S.}~\bibnamefont
  {Zhdanov}}, \bibinfo {author} {\bibfnamefont {H.~M.}\ \bibnamefont
  {Thomas}},\ and\ \bibinfo {author} {\bibfnamefont {G.~E.}\ \bibnamefont
  {Morfill}},\ }\bibfield  {title} {\bibinfo {title} {Highly resolved
  self-excited density waves in a complex plasma},\ }\href
  {https://doi.org/10.1103/PhysRevLett.99.095002} {\bibfield  {journal}
  {\bibinfo  {journal} {Phys. Rev. Lett.}\ }\textbf {\bibinfo {volume} {99}},\
  \bibinfo {pages} {095002} (\bibinfo {year} {2007})}\BibitemShut {NoStop}%
\bibitem [{\citenamefont {Fortov}\ \emph {et~al.}(2003)\citenamefont {Fortov},
  \citenamefont {Vaulina}, \citenamefont {Petrov}, \citenamefont {Molotkov},
  \citenamefont {Chernyshev}, \citenamefont {Lipaev}, \citenamefont {Morfill},
  \citenamefont {Thomas}, \citenamefont {Rothermel}, \citenamefont {Khrapak},
  \citenamefont {Semenov}, \citenamefont {Ivanov}, \citenamefont {Krikalev},\
  and\ \citenamefont {Gidzenko}}]{Fortov2003}%
  \BibitemOpen
  \bibfield  {author} {\bibinfo {author} {\bibfnamefont {V.~E.}\ \bibnamefont
  {Fortov}}, \bibinfo {author} {\bibfnamefont {O.~S.}\ \bibnamefont {Vaulina}},
  \bibinfo {author} {\bibfnamefont {O.~F.}\ \bibnamefont {Petrov}}, \bibinfo
  {author} {\bibfnamefont {V.~I.}\ \bibnamefont {Molotkov}}, \bibinfo {author}
  {\bibfnamefont {A.~V.}\ \bibnamefont {Chernyshev}}, \bibinfo {author}
  {\bibfnamefont {A.~M.}\ \bibnamefont {Lipaev}}, \bibinfo {author}
  {\bibfnamefont {G.}~\bibnamefont {Morfill}}, \bibinfo {author} {\bibfnamefont
  {H.}~\bibnamefont {Thomas}}, \bibinfo {author} {\bibfnamefont
  {H.}~\bibnamefont {Rothermel}}, \bibinfo {author} {\bibfnamefont {S.~A.}\
  \bibnamefont {Khrapak}}, \bibinfo {author} {\bibfnamefont {Y.~P.}\
  \bibnamefont {Semenov}}, \bibinfo {author} {\bibfnamefont {A.~I.}\
  \bibnamefont {Ivanov}}, \bibinfo {author} {\bibfnamefont {S.~K.}\
  \bibnamefont {Krikalev}},\ and\ \bibinfo {author} {\bibfnamefont {Y.~P.}\
  \bibnamefont {Gidzenko}},\ }\bibfield  {title} {\bibinfo {title} {Dynamics of
  macroparticles in a dusty plasma under microgravity conditions (first
  experiments on board the iss)},\ }\href@noop {} {\bibfield  {journal}
  {\bibinfo  {journal} {JETP}\ }\textbf {\bibinfo {volume} {96}},\ \bibinfo
  {pages} {704} (\bibinfo {year} {2003})}\BibitemShut {NoStop}%
\bibitem [{\citenamefont {Goedheer}\ and\ \citenamefont
  {Akdim}(2003)}]{Goedheer2003}%
  \BibitemOpen
  \bibfield  {author} {\bibinfo {author} {\bibfnamefont {W.~J.}\ \bibnamefont
  {Goedheer}}\ and\ \bibinfo {author} {\bibfnamefont {M.~R.}\ \bibnamefont
  {Akdim}},\ }\bibfield  {title} {\bibinfo {title} {Vortices in dust clouds
  under microgravity: A simple explanation},\ }\href
  {https://doi.org/10.1103/PhysRevE.68.045401} {\bibfield  {journal} {\bibinfo
  {journal} {Phys. Rev. E}\ }\textbf {\bibinfo {volume} {68}},\ \bibinfo
  {pages} {045401(R)} (\bibinfo {year} {2003})}\BibitemShut {NoStop}%
\bibitem [{\citenamefont {Schwabe}\ \emph {et~al.}(2014)\citenamefont
  {Schwabe}, \citenamefont {Zhdanov}, \citenamefont {R\"{a}th}, \citenamefont
  {Graves}, \citenamefont {Thomas},\ and\ \citenamefont
  {Morfill}}]{Schwabe2014}%
  \BibitemOpen
  \bibfield  {author} {\bibinfo {author} {\bibfnamefont {M.}~\bibnamefont
  {Schwabe}}, \bibinfo {author} {\bibfnamefont {S.}~\bibnamefont {Zhdanov}},
  \bibinfo {author} {\bibfnamefont {C.}~\bibnamefont {R\"{a}th}}, \bibinfo
  {author} {\bibfnamefont {D.~B.}\ \bibnamefont {Graves}}, \bibinfo {author}
  {\bibfnamefont {H.~M.}\ \bibnamefont {Thomas}},\ and\ \bibinfo {author}
  {\bibfnamefont {G.~E.}\ \bibnamefont {Morfill}},\ }\bibfield  {title}
  {\bibinfo {title} {Collective effects in vortex movement in complex
  plasmas},\ }\href {https://doi.org/10.1103/PhysRevLett.112.115002} {\bibfield
   {journal} {\bibinfo  {journal} {Phys. Rev. Lett.}\ }\textbf {\bibinfo
  {volume} {112}},\ \bibinfo {pages} {115002} (\bibinfo {year}
  {2014})}\BibitemShut {NoStop}%
\bibitem [{\citenamefont {Bockwoldt}\ \emph {et~al.}(2014)\citenamefont
  {Bockwoldt}, \citenamefont {Arp}, \citenamefont {Menzel},\ and\ \citenamefont
  {Piel}}]{Bockwoldt2014}%
  \BibitemOpen
  \bibfield  {author} {\bibinfo {author} {\bibfnamefont {T.}~\bibnamefont
  {Bockwoldt}}, \bibinfo {author} {\bibfnamefont {O.}~\bibnamefont {Arp}},
  \bibinfo {author} {\bibfnamefont {K.~O.}\ \bibnamefont {Menzel}},\ and\
  \bibinfo {author} {\bibfnamefont {A.}~\bibnamefont {Piel}},\ }\bibfield
  {title} {\bibinfo {title} {On the origin of dust vortices in complex plasmas
  under microgravity conditions},\ }\href {https://doi.org/10.1063/1.4897181}
  {\bibfield  {journal} {\bibinfo  {journal} {Phys. Plasmas}\ }\textbf
  {\bibinfo {volume} {21}},\ \bibinfo {pages} {103703} (\bibinfo {year}
  {2014})}\BibitemShut {NoStop}%
\bibitem [{\citenamefont {{Thomas Jr.}}(1999)}]{ThomasJr1999}%
  \BibitemOpen
  \bibfield  {author} {\bibinfo {author} {\bibfnamefont {E.}~\bibnamefont
  {{Thomas Jr.}}},\ }\bibfield  {title} {\bibinfo {title} {Direct measurements
  of two-dimensional velocity profiles in direct current glow discharge dusty
  plasmas},\ }\href {https://doi.org/10.1063/1.873544} {\bibfield  {journal}
  {\bibinfo  {journal} {Phys. Plasmas}\ ,\ \bibinfo {pages} {2672}} (\bibinfo
  {year} {1999})}\BibitemShut {NoStop}%
\bibitem [{\citenamefont {Tsytovich}\ \emph {et~al.}(2004)\citenamefont
  {Tsytovich}, \citenamefont {Vladimirov},\ and\ \citenamefont
  {Morfill}}]{TVM.PRE.2004}%
  \BibitemOpen
  \bibfield  {author} {\bibinfo {author} {\bibfnamefont {V.}~\bibnamefont
  {Tsytovich}}, \bibinfo {author} {\bibfnamefont {S.}~\bibnamefont
  {Vladimirov}},\ and\ \bibinfo {author} {\bibfnamefont {G.}~\bibnamefont
  {Morfill}},\ }\bibfield  {title} {\bibinfo {title} {Theory of dust and
  dust-void structures in the presence of the ion diffusion},\ }\href
  {https://doi.org/10.1103/physreve.70.066408} {\bibfield  {journal} {\bibinfo
  {journal} {Physical Review E}\ }\textbf {\bibinfo {volume} {70}},\ \bibinfo
  {pages} {066408} (\bibinfo {year} {2004})}\BibitemShut {NoStop}%
\bibitem [{\citenamefont {Kananovich}\ and\ \citenamefont
  {Goree}(2020)}]{Kananovich2020a}%
  \BibitemOpen
  \bibfield  {author} {\bibinfo {author} {\bibfnamefont {A.}~\bibnamefont
  {Kananovich}}\ and\ \bibinfo {author} {\bibfnamefont {J.}~\bibnamefont
  {Goree}},\ }\bibfield  {title} {\bibinfo {title} {Shocks propagate in a 2d
  dusty plasma with less attenuation than due to gas friction alone},\ }\href
  {https://doi.org/10.1063/5.0016504} {\bibfield  {journal} {\bibinfo
  {journal} {Physics of Plasmas}\ }\textbf {\bibinfo {volume} {27}},\ \bibinfo
  {pages} {113704} (\bibinfo {year} {2020})}\BibitemShut {NoStop}%
\bibitem [{\citenamefont {Allen}(1992)}]{Allen_1992}%
  \BibitemOpen
  \bibfield  {author} {\bibinfo {author} {\bibfnamefont {J.~E.}\ \bibnamefont
  {Allen}},\ }\bibfield  {title} {\bibinfo {title} {Probe theory - the orbital
  motion approach},\ }\href {https://doi.org/10.1088/0031-8949/45/5/013}
  {\bibfield  {journal} {\bibinfo  {journal} {Physica Scripta}\ }\textbf
  {\bibinfo {volume} {45}},\ \bibinfo {pages} {497} (\bibinfo {year}
  {1992})}\BibitemShut {NoStop}%
\bibitem [{\citenamefont {Kennedy}\ and\ \citenamefont
  {Allen}(2003)}]{kennedy_allen_2003}%
  \BibitemOpen
  \bibfield  {author} {\bibinfo {author} {\bibfnamefont {R.~V.}\ \bibnamefont
  {Kennedy}}\ and\ \bibinfo {author} {\bibfnamefont {J.~E.}\ \bibnamefont
  {Allen}},\ }\bibfield  {title} {\bibinfo {title} {The floating potential of
  spherical probes and dust grains. ii: Orbital motion theory},\ }\href
  {https://doi.org/10.1017/S0022377803002265} {\bibfield  {journal} {\bibinfo
  {journal} {Journal of Plasma Physics}\ }\textbf {\bibinfo {volume} {69}},\
  \bibinfo {pages} {485–506} (\bibinfo {year} {2003})}\BibitemShut {NoStop}%
\bibitem [{\citenamefont {Ratynskaia}\ \emph {et~al.}(2004)\citenamefont
  {Ratynskaia}, \citenamefont {Khrapak}, \citenamefont {Zobnin}, \citenamefont
  {Thoma}, \citenamefont {Kretschmer}, \citenamefont {Usachev}, \citenamefont
  {Yaroshenko}, \citenamefont {Quinn}, \citenamefont {Morfill}, \citenamefont
  {Petrov},\ and\ \citenamefont {Fortov}}]{PhysRevLett.93.085001.Ratynskaia}%
  \BibitemOpen
  \bibfield  {author} {\bibinfo {author} {\bibfnamefont {S.}~\bibnamefont
  {Ratynskaia}}, \bibinfo {author} {\bibfnamefont {S.}~\bibnamefont {Khrapak}},
  \bibinfo {author} {\bibfnamefont {A.}~\bibnamefont {Zobnin}}, \bibinfo
  {author} {\bibfnamefont {M.~H.}\ \bibnamefont {Thoma}}, \bibinfo {author}
  {\bibfnamefont {M.}~\bibnamefont {Kretschmer}}, \bibinfo {author}
  {\bibfnamefont {A.}~\bibnamefont {Usachev}}, \bibinfo {author} {\bibfnamefont
  {V.}~\bibnamefont {Yaroshenko}}, \bibinfo {author} {\bibfnamefont {R.~A.}\
  \bibnamefont {Quinn}}, \bibinfo {author} {\bibfnamefont {G.~E.}\ \bibnamefont
  {Morfill}}, \bibinfo {author} {\bibfnamefont {O.}~\bibnamefont {Petrov}},\
  and\ \bibinfo {author} {\bibfnamefont {V.}~\bibnamefont {Fortov}},\
  }\bibfield  {title} {\bibinfo {title} {Experimental determination of
  dust-particle charge in a discharge plasma at elevated pressures},\ }\href
  {https://doi.org/10.1103/PhysRevLett.93.085001} {\bibfield  {journal}
  {\bibinfo  {journal} {Phys. Rev. Lett.}\ }\textbf {\bibinfo {volume} {93}},\
  \bibinfo {pages} {085001} (\bibinfo {year} {2004})}\BibitemShut {NoStop}%
\bibitem [{\citenamefont {Khrapak}\ \emph {et~al.}(2005)\citenamefont
  {Khrapak}, \citenamefont {Ratynskaia}, \citenamefont {Zobnin}, \citenamefont
  {Usachev}, \citenamefont {Yaroshenko}, \citenamefont {Thoma}, \citenamefont
  {Kretschmer}, \citenamefont {H\"ofner}, \citenamefont {Morfill},
  \citenamefont {Petrov},\ and\ \citenamefont
  {Fortov}}]{PhysRevE.72.016406.Khrapak}%
  \BibitemOpen
  \bibfield  {author} {\bibinfo {author} {\bibfnamefont {S.~A.}\ \bibnamefont
  {Khrapak}}, \bibinfo {author} {\bibfnamefont {S.~V.}\ \bibnamefont
  {Ratynskaia}}, \bibinfo {author} {\bibfnamefont {A.~V.}\ \bibnamefont
  {Zobnin}}, \bibinfo {author} {\bibfnamefont {A.~D.}\ \bibnamefont {Usachev}},
  \bibinfo {author} {\bibfnamefont {V.~V.}\ \bibnamefont {Yaroshenko}},
  \bibinfo {author} {\bibfnamefont {M.~H.}\ \bibnamefont {Thoma}}, \bibinfo
  {author} {\bibfnamefont {M.}~\bibnamefont {Kretschmer}}, \bibinfo {author}
  {\bibfnamefont {H.}~\bibnamefont {H\"ofner}}, \bibinfo {author}
  {\bibfnamefont {G.~E.}\ \bibnamefont {Morfill}}, \bibinfo {author}
  {\bibfnamefont {O.~F.}\ \bibnamefont {Petrov}},\ and\ \bibinfo {author}
  {\bibfnamefont {V.~E.}\ \bibnamefont {Fortov}},\ }\bibfield  {title}
  {\bibinfo {title} {Particle charge in the bulk of gas discharges},\ }\href
  {https://doi.org/10.1103/PhysRevE.72.016406} {\bibfield  {journal} {\bibinfo
  {journal} {Phys. Rev. E}\ }\textbf {\bibinfo {volume} {72}},\ \bibinfo
  {pages} {016406} (\bibinfo {year} {2005})}\BibitemShut {NoStop}%
\bibitem [{\citenamefont {Khrapak}\ \emph
  {et~al.}(2012{\natexlab{b}})\citenamefont {Khrapak}, \citenamefont {Tolias},
  \citenamefont {Ratynskaia}, \citenamefont {Chaudhuri}, \citenamefont
  {Zobnin}, \citenamefont {Usachev}, \citenamefont {Rau}, \citenamefont
  {Thoma}, \citenamefont {Petrov}, \citenamefont {Fortov},\ and\ \citenamefont
  {Morfill}}]{Khrapak_2012}%
  \BibitemOpen
  \bibfield  {author} {\bibinfo {author} {\bibfnamefont {S.~A.}\ \bibnamefont
  {Khrapak}}, \bibinfo {author} {\bibfnamefont {P.}~\bibnamefont {Tolias}},
  \bibinfo {author} {\bibfnamefont {S.}~\bibnamefont {Ratynskaia}}, \bibinfo
  {author} {\bibfnamefont {M.}~\bibnamefont {Chaudhuri}}, \bibinfo {author}
  {\bibfnamefont {A.}~\bibnamefont {Zobnin}}, \bibinfo {author} {\bibfnamefont
  {A.}~\bibnamefont {Usachev}}, \bibinfo {author} {\bibfnamefont
  {C.}~\bibnamefont {Rau}}, \bibinfo {author} {\bibfnamefont {M.~H.}\
  \bibnamefont {Thoma}}, \bibinfo {author} {\bibfnamefont {O.~F.}\ \bibnamefont
  {Petrov}}, \bibinfo {author} {\bibfnamefont {V.~E.}\ \bibnamefont {Fortov}},\
  and\ \bibinfo {author} {\bibfnamefont {G.~E.}\ \bibnamefont {Morfill}},\
  }\bibfield  {title} {\bibinfo {title} {Grain charging in an intermediately
  collisional plasma},\ }\href {https://doi.org/10.1209/0295-5075/97/35001}
  {\bibfield  {journal} {\bibinfo  {journal} {{EPL} (Europhysics Letters)}\
  }\textbf {\bibinfo {volume} {97}},\ \bibinfo {pages} {35001} (\bibinfo {year}
  {2012}{\natexlab{b}})}\BibitemShut {NoStop}%
\bibitem [{\citenamefont {Liu}\ \emph {et~al.}(2003)\citenamefont {Liu},
  \citenamefont {Goree}, \citenamefont {Nosenko},\ and\ \citenamefont
  {Boufendi}}]{liu2003}%
  \BibitemOpen
  \bibfield  {author} {\bibinfo {author} {\bibfnamefont {B.}~\bibnamefont
  {Liu}}, \bibinfo {author} {\bibfnamefont {J.}~\bibnamefont {Goree}}, \bibinfo
  {author} {\bibfnamefont {V.}~\bibnamefont {Nosenko}},\ and\ \bibinfo {author}
  {\bibfnamefont {L.}~\bibnamefont {Boufendi}},\ }\bibfield  {title} {\bibinfo
  {title} {Radiation pressure and gas drag forces on a melamine-formaldehyde
  microsphere in a dusty plasma},\ }\href {https://doi.org/10.1063/1.1526701}
  {\bibfield  {journal} {\bibinfo  {journal} {Phys. Plasmas}\ }\textbf
  {\bibinfo {volume} {10}},\ \bibinfo {pages} {9} (\bibinfo {year}
  {2003})}\BibitemShut {NoStop}%
\end{thebibliography}%






\end{document}